\documentclass[11pt]{article}
\usepackage{axodraw}
\usepackage{epsfig}
\usepackage{amsfonts}
\usepackage{amsmath}
\usepackage{bbm}
\allowdisplaybreaks[1]

\input pix.sty

\renewcommand{\eq}{eq.~}
\renewcommand{\eqs}{eqs.~}
\renewcommand{\se}{sec.~}

\renewcommand{\fig}{fig.~}
\renewcommand{\figs}{figs.~}

\newcommand{\tinymsbar}{{\overline{\mbox{\tiny\rm{MS}}}}}
\newcommand{\Lambdamsbar}{{\Lambda_\tinymsbar}}

\newcommand{\Nf}{N_{\rm f}}
\newcommand{\Nc}{N_{\rm c}}

\newcommand{\Tc}{T_{\rm c}}

\newcommand{\rmO}{{\mathcal{O}}}
\newcommand{\bmu}{\bar\mu}

\newcommand{\CF}{C_F}
\newcommand{\gE}{g_\rmii{E}}

\def\lsi{\raise0.3ex\hbox{$<$\kern-0.75em\raise-1.1ex\hbox{$\sim$}}}
\def\gsi{\raise0.3ex\hbox{$>$\kern-0.75em\raise-1.1ex\hbox{$\sim$}}}
\newcommand{\lsim}{\mathop{\lsi}}

\newcommand{\nF}{n_\rmii{F}}
\newcommand{\nB}{n_\rmii{B}}
 \renewcommand{\nF}[1]{n_\rmii{F{#1}}}
 \renewcommand{\nB}[1]{n_\rmii{B{#1}}}
\newcommand{\rmii}[1]{{\mbox{\tiny\rm{#1}}}}

\newcommand{\Tint}[1]{{\hbox{$\sum$}\!\!\!\!\!\!\!\int\,}_{\!\!\!\!\raise-0.9ex\hbox{$\scriptstyle{#1}$}}}
\newcommand{\Tinti}[1]{{{\Sigma}\!\!\!\!\raise0.3ex\hbox{$\int$}_\rmii{${#1}$}}}

\newcommand{\bi}{\begin{itemize}}
\newcommand{\ei}{\end{itemize}}

\newcommand{\hide}[1]{ }

\def\TAsc(#1,#2)(#3,#4,#5)%
{\SetWidth{2.0}\CArc(#1,#2)(#3,#4,#5)\SetWidth{1.0}}
\def\Lwidth{3}

\def\TAgl(#1,#2)(#3,#4,#5){\SetWidth{2.0}\PhotonArc(#1,#2)(#3,#4,#5){\Lwidth}%
{6.283 #3 mul 360 div #4 #5 sub #4 #5 sub mul sqrt mul Tdensity mul}%
\SetWidth{1.0}}
\def\TLgl(#1,#2)(#3,#4){\SetWidth{2.0}\Photon(#1,#2)(#3,#4){\Lwidth}
{#1 #3 sub #1 #3 sub mul #2 #4 sub #2 #4 sub mul add sqrt Tdensity mul}%
\SetWidth{1.0}}
\newcommand{\piC}[1]{\;\parbox[c]{40pt}{\begin{picture}(120,60)(0,-20)
\SetWidth{1.0}\SetScale{0.35} #1 \end{picture}}\;}
\def\ConnectedA(#1,#2,#3){\piC{#1(60,-15)(75,34,146) #2(60,75)(75,214,326)%
 #3(60,60)(20,190,350)%
 \GBoxc(0,30)(10,10){1} \GBoxc(120,30)(10,10){1}%
  }}
\def\ConnectedB(#1,#2,#3){\piC{#1(60,-15)(75,34,146) #2(60,75)(75,214,326)%
 #3(60,60)(60,0)%
 \GBoxc(0,30)(10,10){1} \GBoxc(120,30)(10,10){1}%
  }}
\def\ConnectedC(#1,#2){\piC{#1(60,-15)(75,34,146) #2(60,75)(75,214,326)%
 \GBoxc(0,30)(10,10){1} \GBoxc(120,30)(10,10){1}%
  }}
\def\ConnectedD(#1,#2){\piC{#1(60,-15)(75,34,146) #2(60,75)(75,214,326)%
 \GBoxc(0,30)(10,10){1} \GBoxc(120,30)(10,10){1}%
 \SetWidth{2.0} 
 \Line(55,55)(65,65)%
 \Line(55,65)(65,55)
  }}

\makeatletter \@addtoreset{equation}{section} \makeatother
\renewcommand{\theequation}{\arabic{section}.\arabic{equation}}
\makeatletter
\renewcommand\section{\@startsection {section}{1}{\z@}%
                                   {-5.5ex \@plus -1ex \@minus -.2ex}
                                   {2.3ex \@plus.2ex}%
                                   {\normalfont\large\bfseries}}
\renewcommand\subsection{\@startsection{subsection}{2}{\z@}%
                                     {-3.25ex\@plus -1ex \@minus -.2ex}%
                                     {1.5ex \@plus .2ex}%
                                     {\normalfont\normalsize\bfseries}}
\renewcommand\thesection {\@arabic\c@section}
\renewcommand\thesubsection   {\thesection.\@arabic\c@subsection}
\renewcommand{\@seccntformat}[1]{%
\csname the#1\endcsname.\hspace{1.0em}}
\makeatother

\begin{document}

\flushbottom

\begin{titlepage}

\begin{flushright}
\vspace*{1cm}
\end{flushright}
\begin{centering}
\vfill

{\Large{\bf
 Massive vector current correlator in thermal QCD
}} 

\vspace{0.8cm}

Y.~Burnier and 
M.~Laine 

\vspace{0.8cm}

{\em
Institute for Theoretical Physics, 
Albert Einstein Center, University of Bern, \\ 
Sidlerstrasse 5, CH-3012 Bern, Switzerland\\}

\vspace*{0.8cm}

\mbox{\bf Abstract}
 
\end{centering}

\vspace*{0.3cm}
 
\noindent
We present an NLO analysis of the massive vector current correlator at
temperatures above a few hundred MeV. The physics of this correlator
originates from a transport peak, related to heavy quark diffusion, and from
the quark-antiquark threshold, related to quarkonium physics. In the bottom
case both can be studied with separate effective theories, but for charm these
may not be accurate, so a study within the full theory is needed. Working in
imaginary time, the NLO correlator can be computed in unresummed perturbation
theory; comparing with lattice data, we find good agreement. Subsequently we
inspect how non-perturbative modifications of the transport peak would affect
the imaginary-time correlator. The massive NLO quark-number susceptibility is 
also contrasted with numerical measurement.

\vfill

 
\vspace*{1cm}
  
\noindent
October
 2012

\vfill

\end{titlepage}

%
\section{Introduction}

Heavy (charm and bottom) quarks are excellent probes for the 
properties of the hot QCD plasma generated in heavy ion collision
experiments. On the theoretical side, 
the existence of a mass scale $M \gg \pi T \gg$~200~MeV renders 
the heavy quarks relatively tractable, permitting for an 
interpolation between the simple
static dynamics of the infinite mass limit
and the high mobility case manifested
by lighter quarks. 
It is particularly fortunate that
two heavy flavours are available, 
offering for a handle
on the functional dependence on $M / \pi T $.
On the experimental side, heavy quarks and quarkonia 
are readily tagged because
of their distinctive leptonic decays. Indeed 
thermal modifications of the bottomonia spectra
were among the first spectacular results
produced by the LHC heavy ion program~\cite{Chatrchyan:2011pe}. 

For the bottom quark case, recent years have seen 
significant progress in theoretical studies of the main 
phenomena involved, namely single quark ``transport''
(diffusion, kinetic equilibration) as well as physics near
the quark-antiquark threshold (quarkonium dissociation, 
chemical equilibration) (cf.\ ref.~\cite{hadron} for a review 
and refs.~\cite{theo_a}--\cite{theo_b} for  
recent contributions). Largely this progress has been 
achieved through the use of modern effective field theory methods (Heavy
Quark Effective Theory, or HQET, for single quarks; Non-Relativistic
QCD, or NRQCD, for quarkonium physics). Once properly formulated
the effective field theory observables can be measured 
non-perturbatively with lattice Monte Carlo methods, and 
indeed first results suggest that these avenues may lead to 
substantial progress~\cite{latt_a}--\cite{latt_b}.

In the charm quark case, however, it is not guaranteed 
that the heavy quark expansion converges fast enough to 
yield quantitatively accurate results for all observables
of interest. On the other hand, 
it no longer appears prohibitively expensive
to treat charm quarks as ``light'' degrees of freedom in lattice QCD. 
Indeed, results have appeared
concerning both thermodynamic quantities~\cite{lat2,milc,buwu} 
and imaginary-time correlators relevant for 
determining dynamical properties of the system~\cite{ding2}
(earlier works can be found in refs.~\cite{olda}--\cite{oldb}
and references therein). 
Yet, a modest scale hierarchy between $\pi T$ and $M$
does exist, and consequently systematic errors, 
both from lattice artifacts and from the unavoidable
analytic continuation~\cite{harvey_rev}, are likely 
to be harder to control than
in the light quark case. In fact, given that systematic
errors related to analytic continuation 
remain substantial even for light quarks~\cite{cond}, 
further crosschecks appear welcome.  

The goal of the current study is to derive results
within next-to-leading order (NLO) perturbation theory which
may help in the analysis of lattice data such as those in 
ref.~\cite{ding2}. In order to allow for a direct appreciation of 
Euclidean measurements, we inspect how specific modifications of 
transport properties and threshold features manifest 
themselves in imaginary time. Ultimately, in accordance
with the philosophy of ref.~\cite{analytic} and practical tests
of refs.~\cite{cond,Bulk_wdep}, the goal would be to 
subtract ``trivial'' ultraviolet features from 
continuum-extrapolated lattice data, in order to allow for 
a model-independent extraction of real-time physics~\cite{cuniberti}.  

On the technical side, the current work represents 
a continuation of our earlier study~\cite{nlo}, in which the 
massive vector current {\em spectral function} was computed at
NLO in the domain $M \gg \pi T$ (keeping only those thermal
effects which are not exponentially suppressed). Here we 
keep the full mass dependence, permitting for an extrapolation 
also to the regime $M \ll \pi T$, as well as contributions
from the transport peak at $\omega \ll 2 M$
which were omitted in ref.~\cite{nlo}. Moreover we work 
directly in {\em imaginary time}, which has the benefit that 
the usual problems of convergence at small $\omega$
are milder. In particular, at NLO infrared safe results can be obtained 
without resummations, similarly to what has previously
been achieved for gluonic observables~\cite{rhoE,Bulk_wdep}. 

The plan of this paper is the following. 
After specifying the observables considered and
discussing the methods employed (\se\ref{se:obs}), 
we outline the qualitative structure 
of our findings in~\se\ref{se:qual}. The detailed analytic
and numerical results of the strict NLO analysis 
comprise~\se\ref{se:res}, whereas in 
secs.~\ref{se:transport} and \ref{se:quarkonium}
the effects of non-perturbative modifications
of the transport peak and quarkonium threshold, respectively, 
are inspected. 
Sec.~\ref{se:concl} presents our conclusions; appendix~A
results for all the ``master'' sum-integrals considered; and 
appendix~B details related to renormalization. 

%
\section{Observables and methods}
\la{se:obs}

%
\subsection{Basic definitions}

The main quantity considered is the vector current correlator
related to a massive flavour. Like in lattice QCD we work in 
Euclidean signature, with the usual thermal boundary conditions
imposed across the time direction. Then the correlator is defined as
\ba
 G^{ }_\rmii{V}(\tau)
 & \equiv & 
 \sum_{\mu = 0}^{d}
 \int_\vec{x} 
 \Bigl\langle 
 (\bar\psi \gamma_\mu \psi) (\tau,\vec{x}) 
 \;
 (\bar\psi \gamma^\mu \psi) (0,\vec{0})
 \Bigr\rangle^{ }_T
  \la{GV_def} \\ 
 & \equiv & 
 G^{ }_{00}(\tau) - G^{ }_{ii}(\tau)
 \;, \quad
 0 < \tau < \frac{1}{T}
 \;, \la{Vii}
\ea
where the Dirac matrices are Minkowskian, and 
a sum over spatial indices is implied. Because of 
current conservation the charge correlator is 
independent of $\tau$, and we denote the 
corresponding ``susceptibility'' by
\be
 \chi^{ }_{ }
 \; \equiv  \; \int_0^\beta \! {\rm d}\tau \, 
  G^{ }_{00}(\tau)
 \; = \; \beta \, G^{ }_{00}(0)
 \;, \quad
 \beta \equiv \frac{1}{T}
 \;.
 \la{chi}
\ee
For future reference let us also record the 
{\em free massless} results for these correlators~\cite{ff}:
\ba
 G_{ii}^\rmi{free}(\tau) & \equiv &   
 2 \Nc T^3 \biggl[ 
  \pi \, (1-2\tau T) \,
  \frac{1 + \cos^2(2\pi\tau T)}{\sin^3(2\pi \tau T)}
 + 
 \frac{2 \cos(2\pi \tau T)}{\sin^2(2\pi \tau T)}
  + \fr16 \biggr]
 \;, \la{GVii} \\ 
 \chi_{ }^\rmi{free} & \equiv & 
 \frac{\Nc T^2}{3}
 \;. \la{chi_free}
\ea
Here $C_A \equiv \Nc = 3$ refers to the number of colours. Later on 
the group theory factor $\CF \equiv (\Nc^2 - 1)/(2\Nc)$
will also appear. Spacetime dimension is denoted by 
$D = d + 1 = 4 - 2 \epsilon$.

At NLO, we find it convenient to 
compute the correlators $G^{ }_\rmii{V}$ and $G^{ }_{00}$.  
The spatial part $G^{ }_{ii}$ 
is then obtained from \eq\nr{Vii}. Verifying explicitly
the $\tau$-independence of $G^{ }_{00}$ provides for a nice
crosscheck of the computation.  

%
\subsection{Wick contractions for the vector current correlator}

The Wick contractions for $G^{ }_\rmii{V}$ are as 
given in ref.~\cite{nlo}. Denoting 
\be
 Q \equiv (\omega_n,\vec{0})
 \;, \quad 
 \Delta^{ }_P \equiv P^2 + M^2
 \;, 
\ee 
the leading-order (LO) vector correlator reads, in momentum space, 
\ba
 \nn[-10mm]
 \ConnectedC(\TAsc,\TAsc) 
 & = & 
 2 C_A \Tint{\{P\}}
 \biggl\{
  \frac{(D-2) Q^2 - 4 M^2}{\Delta^{ }_{P}\Delta^{ }_{P-Q} }
 -   \frac{2(D-2)}{\Delta^{ }_{P} } 
 \biggr\}
 \;. \la{lo} \\[-10mm] \nonumber
\ea
Here $\{ P \}$ denotes fermionic Matsubara momenta, 
and $\Tinti{ \{ P \} } \equiv T\sum_{ \{ p_n \} } \int_\vec{p}$.

At NLO, we have to decide on a meaning
of the renormalized mass. Although conceptually subtle, it is 
technically convenient to employ a pole mass; 
then the bare mass parameter, $M_\rmii{B}^2$, can be expressed as 
$ M_\rmii{B}^2  =   M^2 + \delta M^2$ where at NLO
\ba
 \delta M^2 & = & - g^2 C_F \int_K 
 \biggl\{ 
   (D-2) \biggl[ \frac{1}{K^2} - \frac{1}{\Delta^{ }_{\bar{P} - K} } \biggr]
  + \frac{4 M^2}{K^2 \Delta^{ }_{\bar{P} - K} }
 \biggr\}_{\bar{P}^2 = - M^2}  \hspace*{-5mm} 
 \la{v1} \\ & = & 
 - g^2 C_F \int_\vec{k}
 \frac{1}{2\epsilon_k E_{pk}} 
 \biggl[
  (D-2) (E_{pk} - \epsilon_k )
  + \frac{4 M^2 (\epsilon_k + E_{pk})}
 {(\epsilon_k + E_{pk})^2 - E_p^2}   
 \biggr]
 \la{v2} \\ & = & 
 - \frac{6 g^2 C_F M^2}{(4\pi)^2} 
 \biggl( \frac{1}{\epsilon} + \ln\frac{\bmu^2}{M^2} + \fr43 \biggr) 
 \;.  \la{MB}
\ea
Here $\bmu$ is 
the scale parameter of the $\msbar$ scheme, 
terms of $\rmO(\epsilon)$ were omitted, and 
\be
 \epsilon_k \equiv |\vec{k}|
 \;,  \quad 
 E_p \equiv \sqrt{p^2 + M^2}
 \;, \quad
 E_{pk} \equiv \sqrt{(\vec{p-k})^2 + M^2}
 \;. \la{energies}
\ee
Because of Lorentz
invariance the vector $\bar{P}$ in \eq\nr{v1} can be 
chosen at will, as long as 
we set $p_0 = \pm i E_p$ after carrying out the $K$-integral; this means
that the vector $\vec{p}$ in \eq\nr{v2} is arbitrary. 
The corresponding counterterm contribution reads
\ba
 \nn[-10mm] 
 \ConnectedD(\TAsc,\TAsc) 
 & = & 4 C_A \delta M^2 
 \Tint{\{P\}}
 \biggl\{ 
 \frac{D-2}{\Delta^2_{P} }  -
 \frac{2} {\Delta^{ }_{P} \Delta^{ }_{P-Q} }
 + \frac{4M^2 - (D-2)Q^2}{\Delta^2_{P}\Delta^{ }_{P-Q} }
 \biggr\}
 \;, \la{ct}
\ea
and it is often convenient to identify 
$\vec{p}$ of \eq\nr{v2} as the integration variable
of \eq\nr{ct}.

The ``genuine'' 2-loop graphs amount to
\ba
 \nn[-10mm]
 && \hspace*{-1.5cm}
 \ConnectedA(\TAsc,\TAsc,\TAgl) \; + 
 \ConnectedB(\TAsc,\TAsc,\TLgl) \quad
 = \, 4 g^2 C_A C_F \Tint{K\{P\}} 
 \biggl\{  \nonumber  \\[-5mm] 
 &&  \hspace*{-0.8cm}
 \; \frac{(D-2)^2}{K^2\Delta^2_{P} } - 
 \frac{(D-2)^2}{\Delta^2_{P} \Delta^{ }_{P-K} } -
 \frac{2(D-2)}{K^2\Delta^{ }_{P}\Delta^{ }_{P-K} } +
 \frac{4(D-2)M^2}{K^2\Delta^2_{P}\Delta^{ }_{P-K} }  
 \nn
 && \hspace*{-1cm}
 - \, \frac{2(D-2)}{K^2 \Delta^{ }_{P} \Delta^{ }_{P-Q} }
 + \frac{4 (D-2)M^2 - (D-2)^2Q^2}{K^2 \Delta^2_{P} \Delta^{ }_{P-Q} }
 + \frac{2(D-2)}{K^2 \Delta^{ }_{P-K} \Delta^{ }_{P-Q} }
 \nn
 && \hspace*{-1cm}
 + \, \frac{4(D-2)}{\Delta^{ }_{P} \Delta^{ }_{P-K} \Delta^{ }_{P-Q} }
 - \frac{4 (D-2)M^2 - (D-2)^2Q^2}
        {\Delta^2_{P} \Delta^{ }_{P-K}\Delta^{ }_{P-Q} }
 \nn
 &&  \hspace*{-1cm}
 - \, \frac{16 M^2 + 2 (D-2)^2 K\cdot Q - 4 (D-2) Q^2}
 {K^2 \Delta^{ }_{P} \Delta^{ }_{P-K} \Delta^{ }_{P-Q} } 
 +  \frac{16 M^4 - 4 (D-2)Q^2 M^2}
 {K^2 \Delta^2_{P} \Delta^{ }_{P-K} \Delta^{ }_{P-Q} } 
 \nn
 &&  \hspace*{-1cm} 
 - \, \frac{2(D-4) M^2 + \fr12 (D-2)(8-D) Q^2 + (D-2) K^2}
 {\Delta^{ }_{P} \Delta^{ }_{P-K} \Delta^{ }_{P-Q} \Delta^{ }_{P-K-Q}}
 \nn
 &&  \hspace*{-1cm} 
 +\, \frac{8 M^4 -2 (D-4) M^2 Q^2 - (D-2) Q^4}
 {K^2 \Delta^{ }_{P} \Delta^{ }_{P-K}\Delta^{ }_{P-Q} \Delta^{ }_{P-K-Q} }
 \biggr\}
 \;. \la{nlo}
\ea

%
\subsection{Wick contractions for the susceptibility}

In the case of the zero components, {\em viz.} $G^{ }_{00}$, 
the LO correlator reads, 
in momentum space,  
\ba
 \nn[-10mm]
 \ConnectedC(\TAsc,\TAsc) 
 & = & 
 2 C_A \Tint{\{P\}}
 \biggl\{
   \frac{2}{\Delta^{ }_{P} } 
   - \frac{Q^2 + 4 E_p^2}{\Delta^{ }_{P}\Delta^{ }_{P-Q} }
 \biggr\}
 \;, \la{xlo} \\[-10mm] \nonumber
\ea
whereas the counterterm graph can be expressed as 
\ba
 \nn[-10mm] 
 \ConnectedD(\TAsc,\TAsc) 
 & = & 4 C_A \delta M^2 
 \Tint{\{P\}}
 \biggl\{ 
  \frac{ Q^2 + 4 E_p^2 }{\Delta^2_{P} \Delta^{ }_{P-Q} }
  - \frac{2}{\Delta^{ }_{P} \Delta^{ }_{P-Q} }
  - \frac{1}{\Delta^2_{P} }  
 \biggr\}
 \;. \la{xct} \\[-10mm] \nonumber
\ea
The genuine 2-loop graphs amount to\footnote{%
 The appearance of the ``energy variables'' in the numerators
 implies that there is a certain redundancy in the basis:
 \be
  0   =   
  \Tint{K\{P\}} 
  \biggl\{ 
   \, \frac{2 K\cdot Q}
   {K^2 \Delta^{ }_P\Delta^{ }_{P-K}\Delta^{ }_{P-Q} } 
   -  \frac{Q^2}
  {\Delta^{ }_{P}\Delta^{ }_{P-K}\Delta^{ }_{P-Q} \Delta^{ }_{P-K-Q} } 
  +\, \frac{Q^2 \epsilon_k^2}
  {K^2 \Delta^{ }_{P}\Delta^{ }_{P-K}\Delta^{ }_{P-Q}\Delta^{ }_{P-K-Q} }
  \biggr\}
  \;. \la{redundancy}
 \ee 
 We have used this to eliminate terms containing 
 $Q^2 \epsilon_k^2$ in the numerator.
}
\ba
 \nn[-10mm]
 && \hspace*{-1.5cm}
 \ConnectedA(\TAsc,\TAsc,\TAgl) \; + 
 \ConnectedB(\TAsc,\TAsc,\TLgl) \quad
 = \, 4 g^2 C_A C_F \Tint{K\{P\}} 
 \biggl\{
  \nonumber \\[-5mm] 
 &&  \hspace*{-1cm}
 \; - \, \frac{D-2}{K^2\Delta^2_{P} } + 
 \frac{D-2}{\Delta^2_{P} \Delta^{ }_{P-K} } +
 \frac{2}{K^2\Delta^{ }_{P} \Delta^{ }_{P-K} } -
 \frac{4M^2}{K^2\Delta^2_{P} \Delta^{ }_{P-K} }  
 \nn
 && \hspace*{-1cm}
 - \, \frac{2(D-2)}{K^2 \Delta^{ }_{P} \Delta^{ }_{P-Q} }
 + \frac{(D-2)( 4 E_p^2 +  Q^2) }{K^2 \Delta^2_{P} \Delta^{ }_{P-Q} }
 - \frac{2}{K^2 \Delta^{ }_{P-K} \Delta^{ }_{P-Q} }
 \nn
 && \hspace*{-1cm}
 + \, \frac{4(D-2)}{\Delta^{ }_{P} \Delta^{ }_{P-K} \Delta^{ }_{P-Q} }
 - \frac{ (D-2)( 4 E_p^2 + Q^2) }
        {\Delta^2_{P}\Delta^{ }_{P-K}\Delta^{ }_{P-Q} }
 \nn
 &&  \hspace*{-1cm}
 - \, \frac{16 M^2  - 4 K\cdot Q + 4  Q^2}
 {K^2 \Delta^{ }_{P} \Delta^{ }_{P-K} \Delta^{ }_{P-Q} } 
 +  \frac{4 M^2 ( 4 E_p^2 + Q^2 ) }
 {K^2 \Delta^2_{P} \Delta^{ }_{P-K} \Delta^{ }_{P-Q} } 
 \nn
 &&  \hspace*{-1cm} 
 - \, \frac{(D-2)(E_p^2+E_{pk}^2-\epsilon_k^2 + K^2 + \fr12 Q^2 )-4 M^2 }
 {\Delta^{ }_{P} \Delta^{ }_{P-K} \Delta^{ }_{P-Q} \Delta^{ }_{P-K-Q} }
 \nn
 &&  \hspace*{-1cm} 
 +\, \frac{4M^2(E_p^2+E_{pk}^2-\epsilon_k^2)+ 
  2 Q^2 (E_p^2+E_{pk}^2 + M^2) + Q^4}
 {K^2 \Delta^{ }_{P} \Delta^{ }_{P-K} \Delta^{ }_{P-Q} \Delta^{ }_{P-K-Q} }
 \biggr\}
 \;. \la{xnlo}
\ea

%
\subsection{Matsubara sums and spatial integrals}

The next step is to convert the momentum-space expressions 
to configuration space. 
If we denote the above result 
(\eqs\nr{lo}, \nr{ct}, \nr{nlo})
by $\tilde G^{ }_\rmii{V}(\omega_n)$, 
then the conversion is  obtained as 
\be
 G^{ }_\rmii{V}(\tau) = T \sum_{\omega_n} e^{-i \omega_n \tau}
 \, 
 \tilde G^{ }_\rmii{V}(\omega_n)
 \;, \la{sum}
\ee
and similarly for $G^{ }_{00}$. At NLO 
we are thereby faced with a three-fold Matsubara sum. Making
use of standard techniques, reviewed in some detail in 
ref.~\cite{nlo}, these sums can be carried out in a closed form, 
whereby we are left with integrals over at most two spatial momenta
(i.e.\ two radial directions and one angle). Intermediate
results at this stage are displayed for all individual
master sum-integrals in appendix~A, and for their sums
in appendix~B, in the latter case with the cancellation
of $1/\epsilon$-divergences verified as well.

Two interesting crosschecks are available.
First of all, all $\tau$-dependent terms disappear from $G^{ }_{00}$, 
cf.\ \eq\nr{xNLO_const}. Second, individual parts of the expressions
contain ``contact terms'' $\propto \delta^{ }_\beta(\tau)$, where 
$\delta^{ }_\beta$ denotes the periodic Dirac-delta. These arise
from structures that are independent of $\omega_n$, but also from 
sum-integrals in which $Q^2$ appears in the numerator. It can be verified, 
however, that all contact terms cancel, both at LO and at NLO. 

It remains to carry out the spatial integrals. 
We write 
\be
 \int_\vec{p,k} = \int_{p,k}\int_{z} \;,
\ee
where the normalization of the angular variable 
$z = \cos \theta_\vec{p,k}$ is chosen so that $\int_z 1 = 1$.
Depending on the case, it is convenient to substitute variables as 
\be
 \int_z  \; = \; \frac{1}{2 p k}
 \int_{E_{pk}^-}^{E_{pk}^+} 
 \! {\rm d}E_{pk}^{ } \, E_{pk}^{ }
 \; = \; 
 \frac{1}{2 p k}
 \int_{\epsilon_{pk}^-}^{\epsilon_{pk}^+}
 \! {\rm d}\epsilon_{pk}^{ } \, \epsilon_{pk}^{ }
 \;,
\ee 
where
$
 E_{pk}^\pm \equiv \sqrt{(p\pm k)^2 + M^2}
$, 
$
 \epsilon_{pk}^\pm \equiv |p \pm k|
$.
The angular integrals are doable in most cases. In addition, 
it is also possible to carry out partial integrations with respect 
to the radial directions, which helps to reduce the number of 
independent terms (for the massless case, see ref.~\cite{ns}
for a recent discussion). Closed {\em massless} loop integrals are 
typically solvable, but in general the integrations remain 
to be carried out numerically. Our final expressions are given in 
the next two sections, 
cf.\ \eqs\nr{GVLOtaul}--\nr{GiiLOconst}, 
\nr{susc_final}, \nr{NLO_tau_final}, \nr{NLO_const_final}.  

%
\section{Leading order results and qualitative pattern}
\la{se:qual}

In order to illustrate the qualitative structure of the results, we recall 
in this section the LO expressions for the quantities considered. In general,
two types of contributions appear: those that depend on $\tau$, 
and those that are constant. To display the $\tau$-dependence 
we introduce the periodic dimensionless function
\be
 D_{E_1 \cdots E_k}^{E_{k+1} \cdots E_n}(\tau) 
 \equiv
 \frac{
 e^{ (E_1 + \cdots + E_k)(\beta - \tau) +
     (E_{k+1} + \cdots + E_n)\tau } 
   + 
 e^{ (E_1 + \cdots + E_k)\tau +
 (E_{k+1} + \cdots + E_n)(\beta - \tau) } 
 }{[e^{\beta E_1} \pm 1] \cdots [e^{\beta E_n} \pm 1]}
 \;. \la{D_def}
\ee
In the denominator, the sign is chosen according to whether the 
particle is a boson or a fermion (at NLO, there is only one boson, 
with the on-shell energy denoted by $\epsilon_k$, cf.\ \eq\nr{energies}). 
We also denote $D^{ }_{2E_p} \equiv D^{ }_{E_pE_p}$.
With these conventions, the LO result for $G^{ }_\rmii{V}$ reads
\ba
 \left. 
   G_\rmii{V}^\rmii{LO}(\tau) 
 \right|_\rmii{$\tau$-dep.}
 & = & 
 \left. 
  -  G_{ii}^\rmii{LO}(\tau) 
 \right|_\rmii{$\tau$-dep.}
 \; = \;
 - 2 C_A 
 \int_{p} 
   \Bigl( 2 + \frac{M^2}{E_p^2} \Bigr) D^{ }_{2E_p}(\tau) 
  \;, \la{GVLOtaul}
 \\
 \left. 
   G_\rmii{V}^\rmii{LO}(\tau) 
 \right|_\rmii{const.}
 & = &  - 4 C_A 
 \int_{p} 
 \frac{M^2 T \nF{}'(E_p)}{E_p^2} 
  \;. \la{GVLOconst}
\ea
Here $\nF{}$ denotes the Fermi distribution 
($\nB{}$ denotes the Bose distribution).
The susceptibility only contains a $\tau$-independent part: 
\be
 T \chi^\rmii{LO}_{ } = G_{00}^\rmii{LO} = - 4 C_A \int_{p} T \nF{}'(E_p)
 \;. \la{suscLO}
\ee
Finally, the constant part of the spatial correlator is obtained
through the use of \eq\nr{Vii}:
\be
 \left. 
 G_{ii}^\rmii{LO}(\tau)
 \right|_\rmii{const.}
 =
 - 4 C_A \int_{p} \Bigl( 1 - \frac{M^2}{E_p^2} \Bigr) T \nF{}'(E_p)
 \;. \la{GiiLOconst}
\ee
To our knowledge none of these leading-order expressions can
be integrated in terms of standard elementary functions. 

In terms of the spectral function, {\em viz.}\ $\rho^{ }_{ii}(\omega)$, 
the constant contribution in \eq\nr{GiiLOconst} arises from
(an infinitely narrow) transport peak around $\omega = 0$, 
whereas the ``fast'' $\tau$-dependence in \eq\nr{GVLOtaul}
originates from the quark-antiquark continuum at $|\omega| \ge 2 M$.
Around the middle of the Euclidean time interval, 
$D^{ }_{2E_p}({\beta} / {2}) = -2 T \nF{}'(E_p)$.
Therefore, both terms contribute 
in a comparable manner at large Euclidean time separations, 
even though they originate from completely different types of physics
(see also ref.~\cite{umeda}).  

At NLO, the same three structures appear as at LO: a constant $G^{ }_{00}$, 
as well as a spatial correlator $G^{ }_{ii}$ which has both 
a constant and a $\tau$-dependent part. 
It is generally believed that the perturbative series for the constant 
part of $G^{ }_{ii}$ breaks down at some point and that 
in the full theory, $G^{ }_{ii}$ has no projection 
to the Matsubara zero mode; the reason is that the spatial
components of the vector current do not couple to conserved charges. 
The corresponding ``slow'' $\tau$-dependence then reflects the 
physics of a ``smeared'' transport peak. Yet it remains 
true that single quark physics from the transport peak and 
quark-antiquark physics from the threshold region are
expected to contribute in a comparable manner to the 
Euclidean correlator around $\tau = \beta/2$.
(This is demonstrated explicitly 
in \figs\ref{fig:transport}, \ref{fig:threshold} below.)

%
\section{NLO results}
\la{se:res}

%
\subsection{Susceptibility}

Proceeding to NLO, the result for 
the susceptibility is 
obtained from \eq\nr{xNLO_const} after partial integrations:
\ba
 T \chi^\rmii{NLO}_{ } = 
   G_{00}^\rmii{NLO} 
 & = &
  4 g^2 C_A C_F 
 \int_{p} 
 \frac{T \nF{}'(E_p)}{ p^2  }
 \int_{k}
 \biggl[ 
  \frac{\nB{}(\epsilon_{k})} {\epsilon_{k}^{ } }
    + 
  \frac{\nF{}(E^{ }_{k})}{E_{k}^{ }}
  \biggl( 1 - \frac{M^2}{k^2} \biggr)
 \biggr]
 \;.  \la{susc_final}
\ea
This can be shown to agree
with what can be extracted from ref.~\cite{phenEOS} as 
$T \partial_\mu^2 \left. p(T,\mu) \right|_{\mu = 0}$.
(The substance of the information is already there in ref.~\cite{jik}.)
The massless loop evaluates to 
$
 \int_k \nB{}(\epsilon_{k}) / \epsilon_{k}^{ } = T^2/12
$, 
and in the limit $M \gg \pi T$
its effect can be interpreted as 
an effective mass correction
to the LO result of \eq\nr{suscLO},
$
 M^2 \to M^2 + {g^2 T^2 \CF} / {6}
$~\cite{dhr}.
On the other hand, 
in the massless limit $M\ll \pi T$
\eq\nr{susc_final} reduces to the 
well-known correction~(cf.\ \cite{av} and references therein)
\be
  G_{00}^\rmii{NLO} 
 \; \stackrel{M \ll \pi T}{=} \;
   - \frac{ g^2 \Nc \CF T^3 }{ 8 \pi^2 } + \rmO(g^3)
 \;.
\ee

A numerical evaluation of \eqs\nr{suscLO}, \nr{susc_final}
is shown in \fig\ref{fig:susc}. For the gauge coupling we have 
inserted the 2-loop value for the thermal coupling $\gE^2 / T$ from 
ref.~\cite{gE2}, and for comparison with quenched 
lattice data from refs.~\cite{ding1,ding2} we assume 
that $\Tc/\Lambdamsbar \simeq 1.25$, $T/\Tc \simeq 1.45$.
In the massive case, where no continuum extrapolation 
is available in ref.~\cite{ding2}, we choose to compare with results
obtained on a lattice $128^3\times 48$. The quark mass cited, 
$m_\rmii{$\msbar$}(m_c) = 1.094(1)$~GeV is phenomenologically 
on the low side~\cite{pdg}; 
based on \eq\nr{MB}, {\em viz.}
\be
 M \simeq 
 m_\rmii{$\msbar$}(m_c)
 \biggl\{ 1  + \frac{4 g^2_\rmii{$\msbar$}(m_c) C_F}{(4\pi)^2} 
 + \rmO(g^4)
 \biggr\}
 \;, 
\ee 
we get $M\simeq 1.3$~GeV but assign a large error to this number. 
Ref.~\cite{ding2} also cites 
$\Tc/\sqrt{\sigma} = 0.630(5)$, 
$\sqrt{\sigma} = 428$~MeV, 
so we estimate $M/T \simeq 3.3\pm 0.5$, 
but with substantial systematic uncertainties, from  
lattice artifacts, string tension measurement~\cite{mt}, 
quenching, as well as perturbative input.  
With this in mind, the excellent agreement seen 
in \fig\ref{fig:susc} is remarkable, and supports
the long-held belief that all quarks, and heavy
quarks in particular, are well described by the weak-coupling
expansion at surprisingly low temperatures. 
(In the massless case the good agreement is consistent with the recent
unquenched study of ref.~\cite{susc}, in which a similar ``dimensionally 
reduced'' resummation scheme was used as here~\cite{phenEOS}, however 
an almost {\em perfect} match already at NLO may be somewhat coincidental.)

\begin{figure}[t]


\centerline{%
 \epsfysize=7.5cm\epsfbox{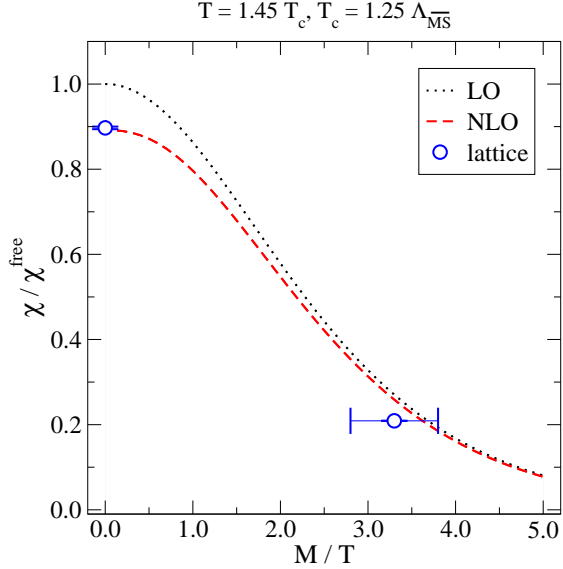}%
}

\caption[a]{\small
The quark-number susceptibility, 
\eqs\nr{suscLO}, \nr{susc_final}, 
normalized to the free result from \eq\nr{chi_free}. 
The lattice result at $M=0$ comes from ref.~\cite{ding1}, 
that at $M > 0$ from ref.~\cite{ding2}; both are quenched 
($\Nf = 0$) and only the former represents the continuum limit.
The uncertainty of $M/T$ is our estimate (cf.\ the text).  
Lattice results for charm  
by other groups can be found in refs.~\cite{milc,buwu}.
}

\la{fig:susc}
\end{figure}

%
\subsection{Vector current correlator}

The vector current correlator is obtained from 
\eqs\nr{NLO_tau}, \nr{NLO_const} after partial integrations.
Its $\tau$-dependent part reads 
\ba
 & & \hspace*{-1cm}
 \frac{ 
  \left. G_\rmii{V}^\rmii{NLO} \right|_\rmii{$\tau$-dep.}
  }{ 
  4 g^2 C_A C_F 
  } = 
 \int_{p}  
 \frac{D^{ }_{2E_p}(\tau)}{4\pi^2} \biggl[ 
 \biggl( 3 + \frac{M^2}{E_p^2} \biggr)
 \biggl( 1 - \frac{p}{2 E_p} \ln \frac{E_p + p}{E_p - p}\biggr)
 - 1
 - \biggl( 2 + \frac{M^2}{E_p^2} \biggr) 
   \int_0^\infty \!\!\! {\rm d}k\, \frac{\theta(k)}{k}
\biggr]
 \nn
 &  + & \!\! \int_{p,k} \mathbbm{P} \Biggl\{ 
 \int_{z} 
 \frac{D^{ }_{\epsilon_k E_p E_{pk}}(\tau)}{\epsilon_k E_p E_{pk}
 \Delta_{+-}\Delta_{-+} }
 \biggl[ - 2 E_p^2 
  + M^2 \biggl( 
  \frac{\Delta_{--}}{\Delta_{++}}
 + \frac{2\epsilon_k^2}{\Delta_{+-}\Delta_{-+}}
 \biggr)
 + \frac{4\epsilon_k^2 M^4}{\Delta_{++}^2 \Delta_{+-}\Delta_{-+}}
 \biggr]
 \nn 
 &  & \qquad + \, 
 \int_{z} 
 \frac{D^{\epsilon_k}_{E_p E_{pk}}(\tau)}{\epsilon_k E_p E_{pk}
 \Delta_{+-}\Delta_{-+}
 }
 \biggl[ - 2 E_p^2 
  + M^2 \biggl( 
  \frac{\Delta_{++}}{\Delta_{--}}
 + \frac{2\epsilon_k^2}{\Delta_{+-}\Delta_{-+}}
 \biggr)
 + \frac{4\epsilon_k^2 M^4}{\Delta_{--}^2 \Delta_{+-}\Delta_{-+}}
 \biggr]
 \nn 
 &  & \qquad + \, 
 \int_{z} 
 \frac{2 D^{E_p}_{\epsilon_k E_{pk}}(\tau)}{\epsilon_k E_p E_{pk}
  \Delta_{++}\Delta_{--}
 }
 \biggl[ 
 E_p^2 + E_{pk}^2
  - M^2 \biggl( 
  \frac{\Delta_{+-}}{\Delta_{-+}}
 + \frac{2\epsilon_k^2}{\Delta_{++}\Delta_{--}}
 \biggr)
 - \frac{4\epsilon_k^2 M^4}{\Delta_{-+}^2 \Delta_{++}\Delta_{--}}
 \biggr]
 \nn 
 & + &  \frac{ D^{ }_{2E_p}(\tau)}{ 2\epsilon_k^3} 
 \biggl(2  + \frac{M^2}{E_p^2} \biggr) 
  \, \biggl[ \;
 -1 + \frac{
    E_p^2 (E_{pk}^+ - E_{pk}^-)
     -
    p\epsilon_k (E_{pk}^+ + E_{pk}^-)
    }{2p(E_p^2 - \epsilon_k^2)}
 + \frac{\epsilon_k^2 M^2 (E_{pk}^+ - E_{pk}^-) }
    {p (E_p^2 - \epsilon_k^2) E_{pk}^+  E_{pk}^- }
 \nn &   & 
 \qquad
  \; + \,
 \frac{ 2 E_p^2  - M^2 }{2p E_p}
 \biggl(  
     \ln\biggl| \frac{(E_p + p)(2p-\epsilon_k)}
                     {(E_p - p)(2p+\epsilon_k)}
        \biggr|
    +\ln\biggl| \frac{1 - {\epsilon_k^2} / {(E_p + E_{pk}^-)^2}}
                 {1 - {\epsilon_k^2} / {(E_p + E_{pk}^+)^2}}
        \biggr|
 \biggr) \;
 + \theta(k) 
 \biggr]
 \nn 
 &  + &
 \frac{D^{ }_{2E_p}(\tau)  \nB{}(\epsilon_k)}{\epsilon_k}
  \biggl[ \;
     \frac{3 M^2}{2 p^2 E_p^2} +   
     \frac{1}{p E_p} \ln\frac{E_p + p}{E_p - p}
  \nn & & \qquad 
   + \, 
   \frac{1}{\epsilon_k^2} \biggl( 2 + \frac{M^2}{E_p^2} \biggr)
   \biggl( - 1 +  \frac{ 2 E_p^2  - M^2 }{2p E_p}
   \ln\frac{E_p + p}{E_p - p}
   \biggr) 
 \; \biggr]
 \nn &  + &    
 \frac{D^{ }_{2E_p}(\tau)  \nF{}(E_{k})}{E_{k}} \, 
 \biggl[ \;
  \frac{3M^2}{2p^2E_p^2}
 + 
   \frac{10 p^4 + 16 p^2 M^2 + 3 M^4}{2({p^2-k^2})p^2E_p^2}
 + \frac{M^2}{p k E_p^2} \ln \biggl| \frac{p+k}{p-k} \biggr|
 \nn &   & 
 \qquad
  - \, 
 \frac{2 E_p^4 + 2 E_k^2 E_p^2 - M^4}{2 p k (E_p - E_k) E_p^3}
 \biggl( 
   \ln \biggl| \frac{p+k}{p-k} \biggr| + 
  \frac{E_k}{E_p + E_k} 
   \ln \biggl| \frac{M^2 + E_p E_k + p k}{M^2 + E_p E_k - p k} \biggr|
 \biggr)
 \; \biggr]
 \Biggr\} 
 \;, \la{NLO_tau_final}
\ea
where $\mathbbm{P}$ denotes a principal value, 
and $\Delta^{ }_{++}$ etc are defined in \eq\nr{def_Delta}. 
The constant contribution reads 
\ba
 \frac{ 
  \left. G_\rmii{V}^\rmii{NLO} \right|_\rmii{const.}
  }{ 
  4 g^2 C_A C_F 
  } 
 & = & \int_{p}  T \nF{}'(E_p)\int_{k} \biggl\{ 
  \frac{\nB{}(\epsilon_k)}{\epsilon_k} \biggl[ 
   \frac{1}{p^2} - \frac{3}{E_p^2}
 \biggr]
 \nn & + &  
 \frac{\nF{}(E_{k})}{E_{k}} \biggl[ 
   \frac{1}{p^2} - \frac{3}{E_p^2}
  - \frac{M^2}{p^2k^2} - \frac{M^2}{E_p^2E_k^2}
  + \frac{M^2(4 E_k^2 - M^2)}{2 p k E_p^2 E_k^2}
    \ln \biggl| \frac{p+k}{p-k} \biggr|  
 \biggr]
 \biggr\} \;. \hspace*{8mm}
  \la{NLO_const_final}
\ea
The spatial correlator subsequently follows from \eq\nr{Vii}, 
with the susceptibility inserted from \eq\nr{susc_final}.

Although the expression in
\eq\nr{NLO_tau_final} is finite, its numerical evaluation 
is non-trivial. There are at least three separate challenges: 
at small $k$ various parts of the expression are divergent, and
care must be taken in order 
to avoid significance loss in their cancellation;
at $k=p$ there is a pole which 
is defined in the sense of a principal value; and at large $k$ there 
is a vacuum part which decreases only slowly (although it is 
integrable). For the reader's benefit, let us briefly specify
how we have dealt with these challenges. 

\begin{itemize}

\item
The small-$k$ divergence originates from  
the terms integrated over $z$ in \eq\nr{NLO_tau_final} and 
from terms 
where the integral had already been carried out. 
For the latter type the divergent part reads
\be
 \frac{ D^{ }_{2E_p}(\tau)}{ \epsilon_k^3} 
 \biggl(2  + \frac{M^2}{E_p^2} \biggr) 
  \, \biggl( \;
 -1 + 
 \frac{ 2 E_p^2  - M^2 }{2p E_p}
     \ln\frac{E_p + p}
                     {E_p - p}
 \;
 \biggr)
 \biggl( \; 
   \fr12 + \nB{}(\epsilon_k)
 \, \biggr)
 \;,
\ee
containing both vacuum and Bose-enhanced structures. 
The integral $\int_{z}$ needs to be carried out
precisely enough such that the cancellation takes duly place.

\item
The principal value integration can be handled for instance by 
reflecting the range $p > k$ into the range $p < k$:
\be
 \int_0^\infty \! {\rm d}p \, p^2\, \phi(p) = 
 k^3 \int_0^1 \! {\rm d}x 
 \Bigl[\,
   x^2 \, \phi(k x) + \frac{1}{x^4} \phi\Bigl( \frac{k}{x} \Bigr) 
 \,\Bigr]
 \;.
\ee
Here $\phi$ has to be evaluated precisely enough 
for cancellations at $x = 1$ to take place. 

\item
A possible way to accelerate the convergence at large $k$ 
is with the help of the function $\theta(k)$
in \eq\nr{NLO_tau_final}. (Note that a power tail only appears 
in the vacuum part.) The simplest subtraction removes
just the leading asymptotic behaviour $- 3 E_p^2/k^2$, e.g.\  
\be
 \theta(k) 
 \equiv
 \frac{3 E_p^2 \Theta(k-k_\rmii{min})}{k^2 + \lambda^2}
 \;, \quad
  \int_0^\infty \!\! {\rm d}k\, \frac{\theta(k)}{k}
 = 
  \frac{3 E_p^2  }{2\lambda^2}
  \, \ln \Bigl( 1 + \frac{\lambda^2}{k_\rmii{min}^2} \Bigr)
 \;, 
\ee 
but of course more refined choices can be envisaged.
(By replacing $\lambda$ by a ``gluon mass'' it would be possible
to take $k_\rmi{min}\to 0$ and still carry out 
$\int_0^\infty \! {\rm d}k \,\theta(k) / k$ exactly, 
cf.\ \eqs\nr{triangle1}, \nr{triangle2}, but the price to pay is 
that then the small-$k$ range has more structure 
than before, with a would-be divergence only cut off at $k \lsim \lambda$.) 

\end{itemize}

\begin{figure}[t]


\centerline{%
 \epsfysize=7.5cm\epsfbox{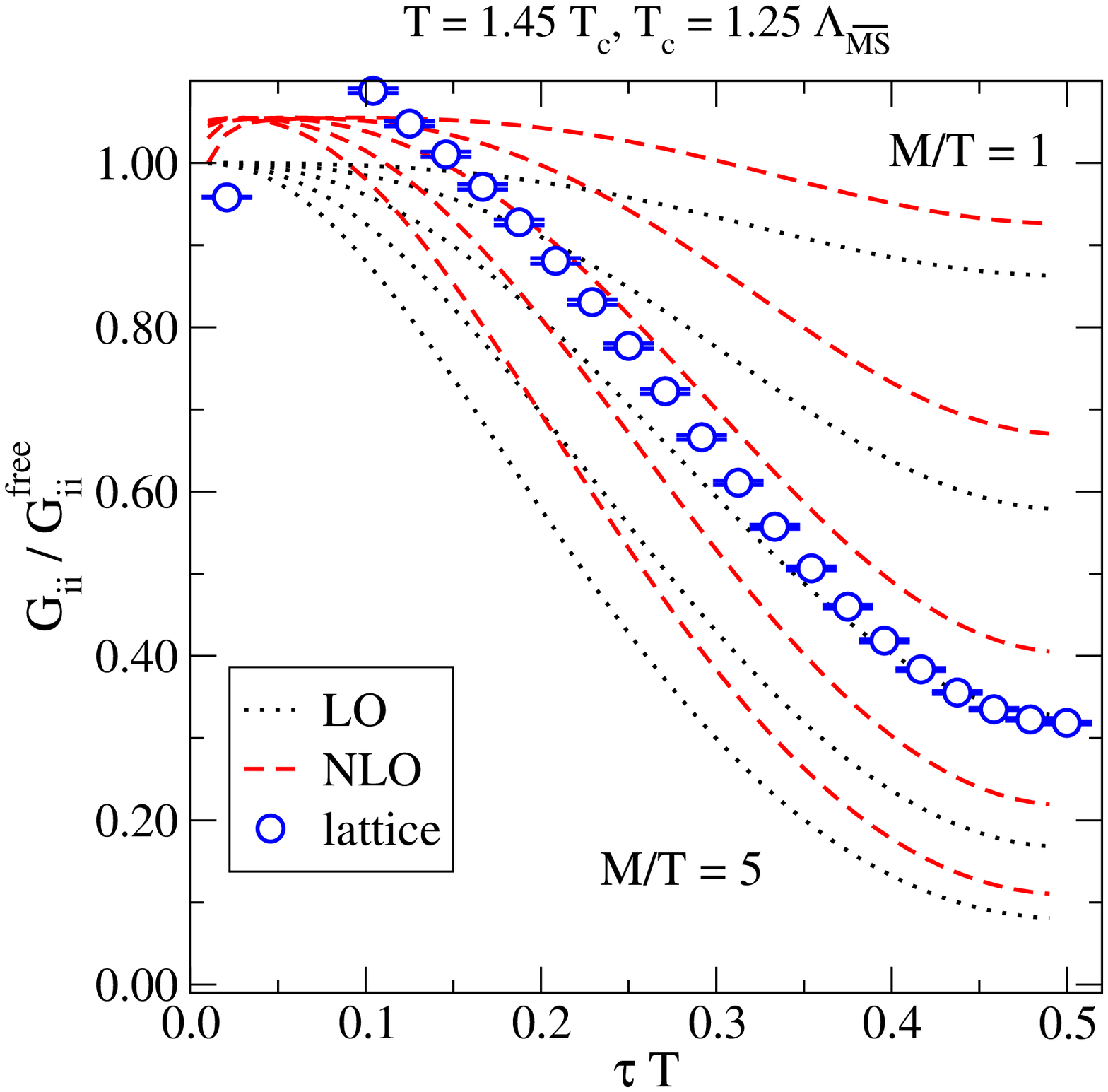}%
~~~\epsfysize=7.5cm\epsfbox{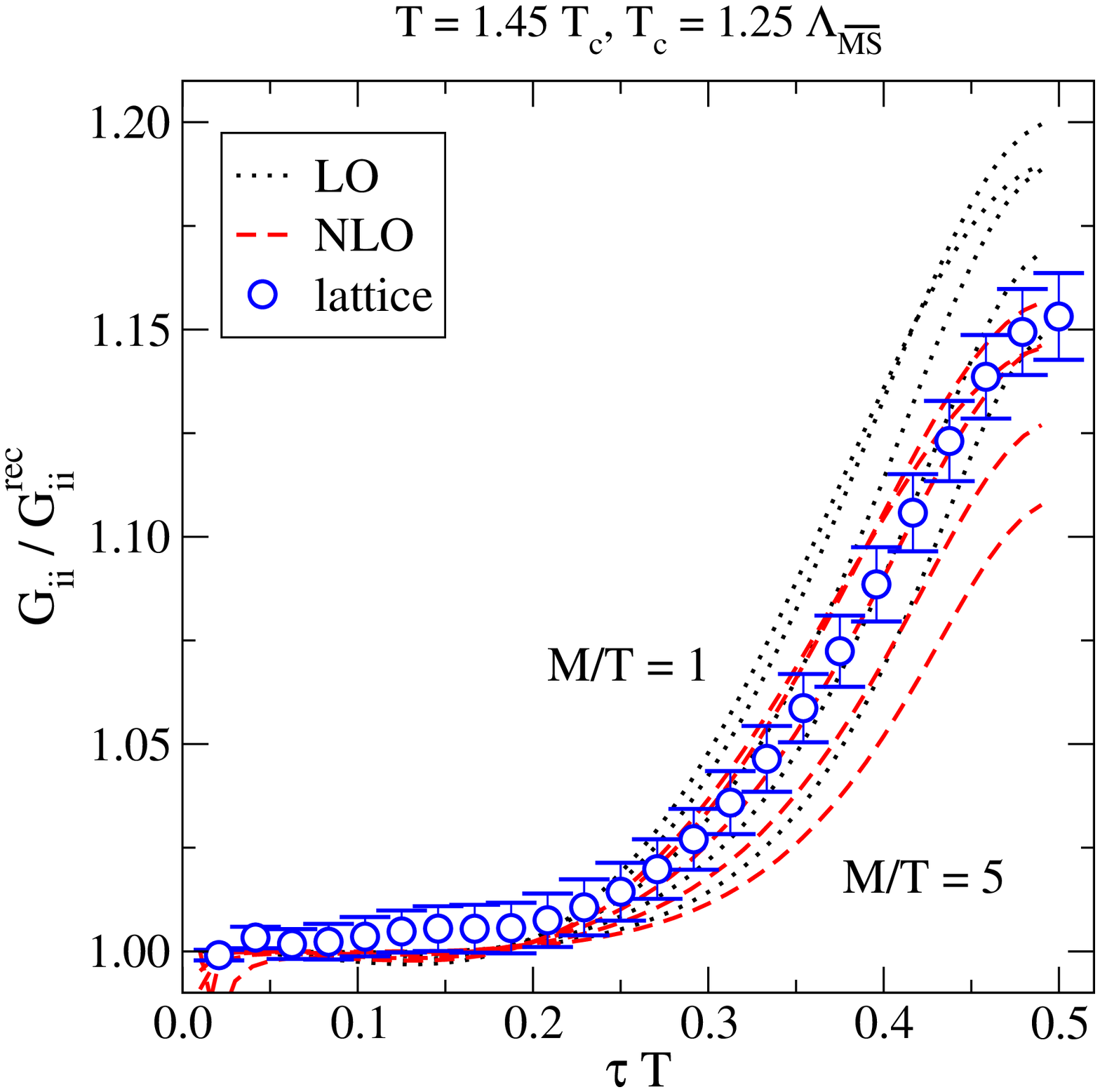}
}

\caption[a]{\small
Left: The vector correlator, 
normalized to the free result from \eq\nr{GVii}, 
for $M/T = 1.0,2.0,3.0,4.0,5.0$ (top to bottom). 
Right: The vector correlator, normalized 
to the ``reconstructed'' result from \eqs\nr{full_vac}, \nr{relation}, 
for the same masses (left to right). 
The lattice results are from ref.~\cite{ding2}; they are  
quenched ($\Nf = 0$) and do not contain a continuum extrapolation
($N_\tau = 48$, $N_s = 128$).
}

\la{fig:Gii}
\end{figure}

For a transparent representation of the numerical results, 
we consider two different normalizations. A simple and theoretically
clean possibility is to normalize
the results to the {\em free  massless} expression from \eq\nr{GVii}.
Another reference point is to make use of the full NLO
spectral function in vacuum~\cite{old}--\cite{db}:
\ba
 \rho^\rmi{vac}_\rmii{V}(\omega) 
 \!\!\! & = & \!\!\!
 - \theta(\omega - 2 M)   
  \frac{C_A (\omega^2 - 4 M^2)^{\fr12}(\omega^2 + 2 M^2)}{4\pi\omega}
 + \theta(\omega - 2 M) 
 \frac{8 g^2 C_A C_F}{(4\pi)^3 \omega^2} 
 \biggl\{ \nn & & 
 (4 M^4 - \omega^4)  L_2 \biggl( \frac{\omega - \sqrt{\omega^2 - 4 M^2}}
 {\omega + \sqrt{\omega^2 - 4 M^2}} \biggr)
 + (7 M^4 + 2 M^2\omega^2 - 3 \omega^4)
  \,\mathrm{acosh} \biggl( \frac{\omega}{2 M} \biggr)
 \nn &  & 
 + \omega (\omega^2 - 4 M^2)^{\fr12}
 \biggl[
   (\omega^2 + 2 M^2) \ln \frac{\omega (\omega^2 - 4 M^2)}{M^3}
  -\fr38 (\omega^2 + 6 M^2) 
 \biggr]
 \biggr\} + \rmO(g^4) \;, 
  \nn \la{full_vac}
\ea
where the function $L_2$ is defined as
\be
 L_2(x) \; \equiv \; 
 4 \, \mathrm{Li}_2 (x) + 2 \, \mathrm{Li}_2(-x)
 + [2 \ln(1-x) + \ln(1+x)] \ln x
 \;. \la{L2} 
\ee
There is no transport peak in the vacuum expression, 
and recalling \eq\nr{Vii}, the corresponding spatial correlator 
can be obtained through 
\be
  G^\rmi{rec}_{ii}(\tau) \equiv
 \int_0^\infty
 \frac{{\rm d}\omega}{\pi} ( - \rho^\rmi{vac}_{V} )(\omega)
 \frac{\cosh \left(\frac{\beta}{2} - \tau\right)\omega}
 {\sinh\frac{\beta \omega}{2}} 
 \;. \la{relation}
\ee
A normalization with respect to a similar ``reconstructed'' correlator
$
  G^\rmi{rec}_{ii}(\tau)
$
has been used in ref.~\cite{ding2} (see also ref.~\cite{Meyer:2010ii}), 
and may be useful for phenomenological purposes, although from the 
theoretical perspective it induces new systematic uncertainties. 

In \fig\ref{fig:Gii} we show our results in both normalizations, 
compared with lattice data from ref.~\cite{ding2}. (The gauge coupling 
has been fixed as explained in connection with \fig\ref{fig:susc}; 
at very small
$\tau$ a different running would appear reasonable 
but in the absence of an NNLO 
computation and continuum-extrapolated lattice data, we stick to the 
simplest choice in the following.)
Like in \fig\ref{fig:susc}, an excellent agreement is found at 
large time separations, if a value $M/T \simeq 3.5$ is assumed. 
(The discrepancy in \fig\ref{fig:Gii}(left) at small $\tau$ is 
probably due to the missing continuum extrapolation.)

%
\section{Modification of the transport peak}
\la{se:transport}

As mentioned in \se\ref{se:qual}, the correlator $G^{ }_{ii}$
has a constant ($\tau$-independent) part at LO and at NLO, but
within the full dynamics
this is expected to turn into a slowly evolving function. 
The purpose of this section is to estimate how
precisely Euclidean data should be measured in order to 
resolve the slow time dependence. 

In order to reach this goal, we model the transport
peak through a Lorentzian shape, 
\be
 \rho^\rmii{(L)}_{ii}(\omega) \; \equiv \; 3 D \chi^{ }_{ } \, 
 \frac{\omega \eta^2}{\omega^2 + \eta^2} 
 \frac{1}{ \cosh(\frac{\omega}{2\pi T}) }
 \;. \la{Lorentz}
\ee
Here $D$ corresponds to the heavy flavour diffusion coefficient. 
The Lorentzian shape can be correct only at small frequencies, 
$|\omega| \ll \pi T$, cf.\ e.g.\ ref.~\cite{amr}; we have chosen
to cut it off at large frequencies through the same recipe that
has been used in the massless case~\cite{cond}.
The susceptibility $\chi$ is fixed according to ref.~\cite{ding2}, 
$\chi/T^2 = 0.20894(1)$. We then vary $D$, in each case tuning 
the other parameter $\eta$ so as to keep the area under 
the transport peak fixed at the value predicted by the NLO expression, 
\eqs\nr{GiiLOconst}, \nr{susc_final}, \nr{NLO_const_final}, \nr{Vii}:
\be
 \frac{1}{\beta}
 \int_0^\beta \! {\rm d}\tau \, G_{ii}^\rmii{(L)}(\tau) 
 \; = \; 
 \int_0^\infty
 \frac{{\rm d}\omega}{\pi}  \frac{ 2 \rho^\rmii{(L)}_{ii}(\omega)}
 { \beta \omega } 
 \; \equiv \; 
 \left[ 
 G_{ii}^\rmii{LO} 
 + 
 G_{ii}^\rmii{NLO} 
 \right]_\rmii{const.}
 \;. \la{constraint} 
\ee
Subsequently the ``correct'' $G^{ }_{ii}(\tau)$ is obtained 
by replacing the part
$
 \left[ 
 G_{ii}^\rmii{LO} 
 + 
 G_{ii}^\rmii{NLO} 
 \right]_\rmii{const.}
$
through a $\tau$-dependent function, 
$
 G_{ii}^\rmii{(L)}(\tau)
$,  
determined by
$
 \rho^\rmii{(L)}_{ii}
$
via \eq\nr{relation}. 
As a guideline, we recall that 
$2\pi T D  \simeq 3 - 5$ may be expected
according to refs.~\cite{Francis:2011gc,Banerjee:2011ra};
that ref.~\cite{ding2} found even
$2\pi T D  \sim 2$; and that in the massless case values 
down to $2\pi T D  \sim 1$ appear possible~\cite{cond}. 

\begin{figure*}


\centerline{%
 \epsfysize=7.5cm\epsfbox{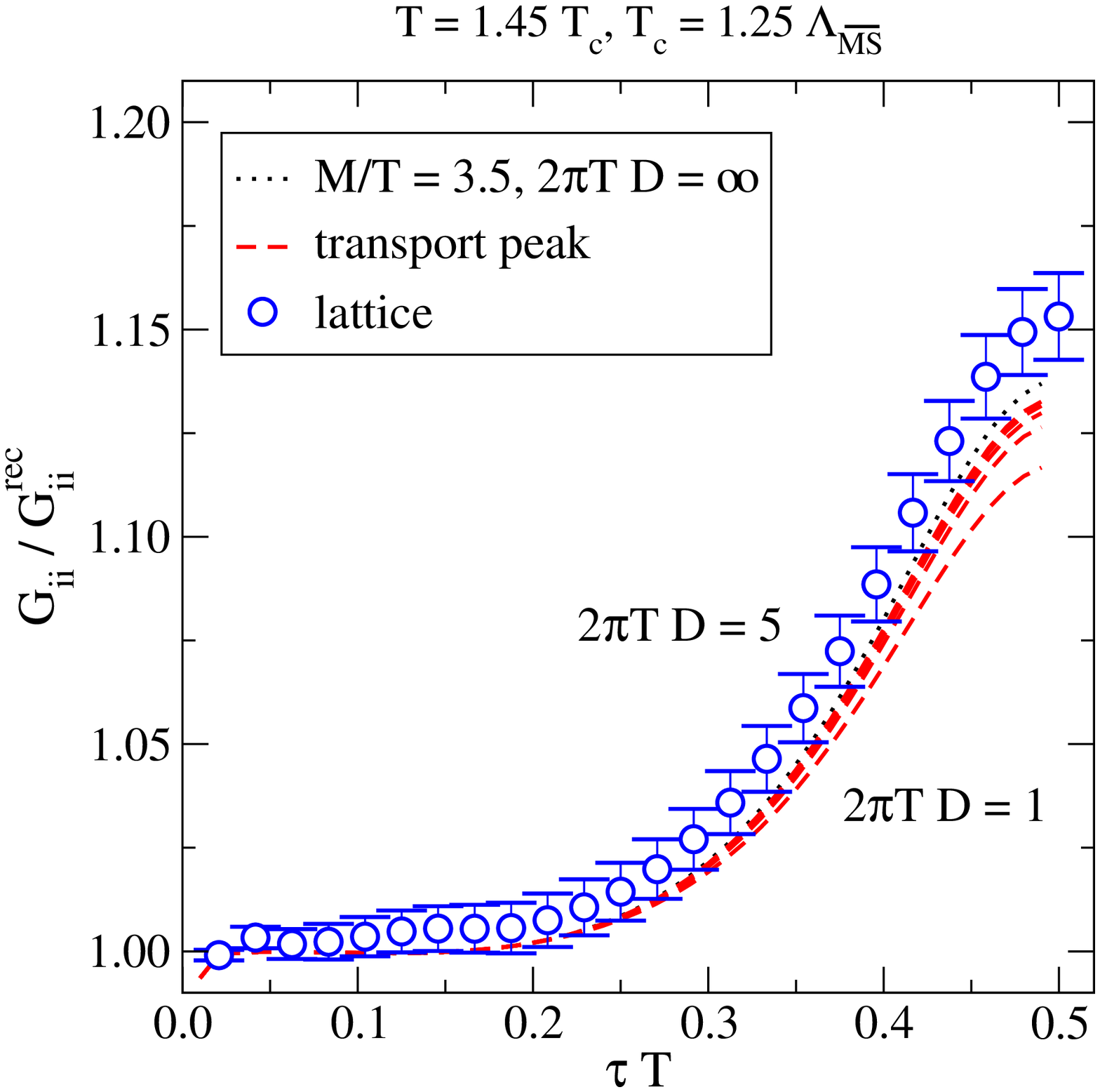}
~~~\epsfysize=7.5cm\epsfbox{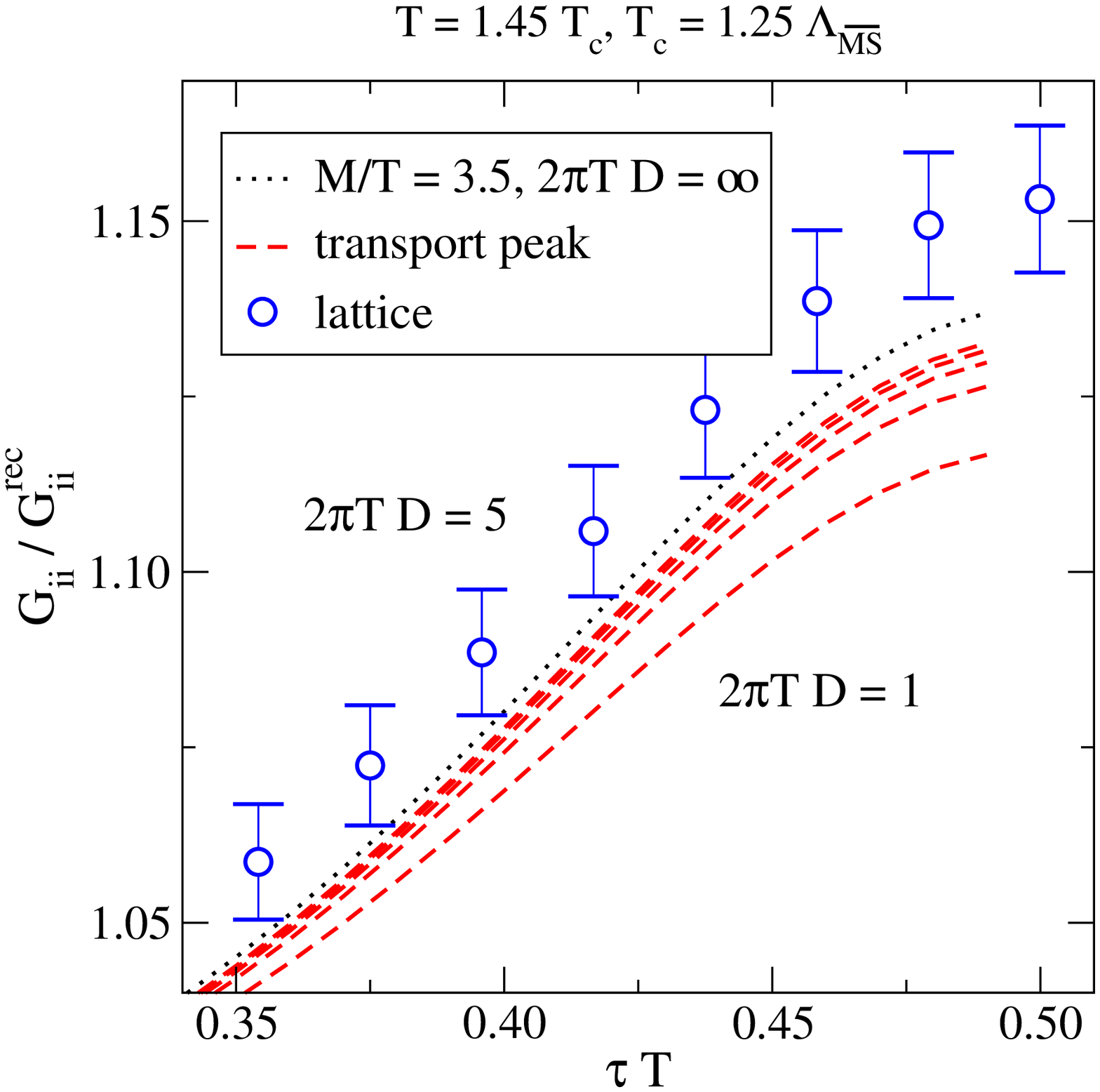}%
}

\caption[a]{\small
Left: The effect of a modified transport peak, for $M/T = 3.5$ and 
$2\pi T D = 1.0$, $2.0$, $3.0$, $4.0$, $5.0$ (cf.\ \eq\nr{Lorentz}). 
Right: A magnification of the large-$\tau$ region. 
It is clear that a high 
precision is needed for resolving the diffusion coefficient from 
the massive vector current correlator. (The curves could
be put on top of lattice data through a minor change of $M/T$, 
but we have refrained from doing this
in the absence of a continuum extrapolation
of the reconstructed correlator.)}

\la{fig:transport}
\end{figure*}

Results are shown in \fig\ref{fig:transport}. 
We observe that if $2\pi T D \sim 1 - 2$, then there is 
hope of resolving it with high enough precision of 
the lattice data 
(it appears that statistical errors should
probably be reduced to 20\% of the current ones to be sensitive
to the features of the transport peak; obviously improvements in 
statistical accuracy need to be accompanied by a corresponding
decrease in systematic uncertainties). 
In the case $2\pi T D \sim 3 - 5$ 
a reliable determination from the massive vector current
correlator appears challenging.  

%
\section{Modification of the threshold region}
\la{se:quarkonium}

The second qualitative structure affecting the massive vector
current correlator, the quarkonium threshold region inducing a 
``fast'' time dependence from the energy scale $\sim 2 M$, is also 
expected to undergo drastic changes in the interacting theory. 
At low temperatures, the spectral function 
is characterized by quarkonium resonances; at high temperatures, 
these are expected to move, broaden, and eventually dissolve 
into a mere threshold enhancement. 

\begin{figure*}


\centerline{%
 \epsfysize=7.5cm\epsfbox{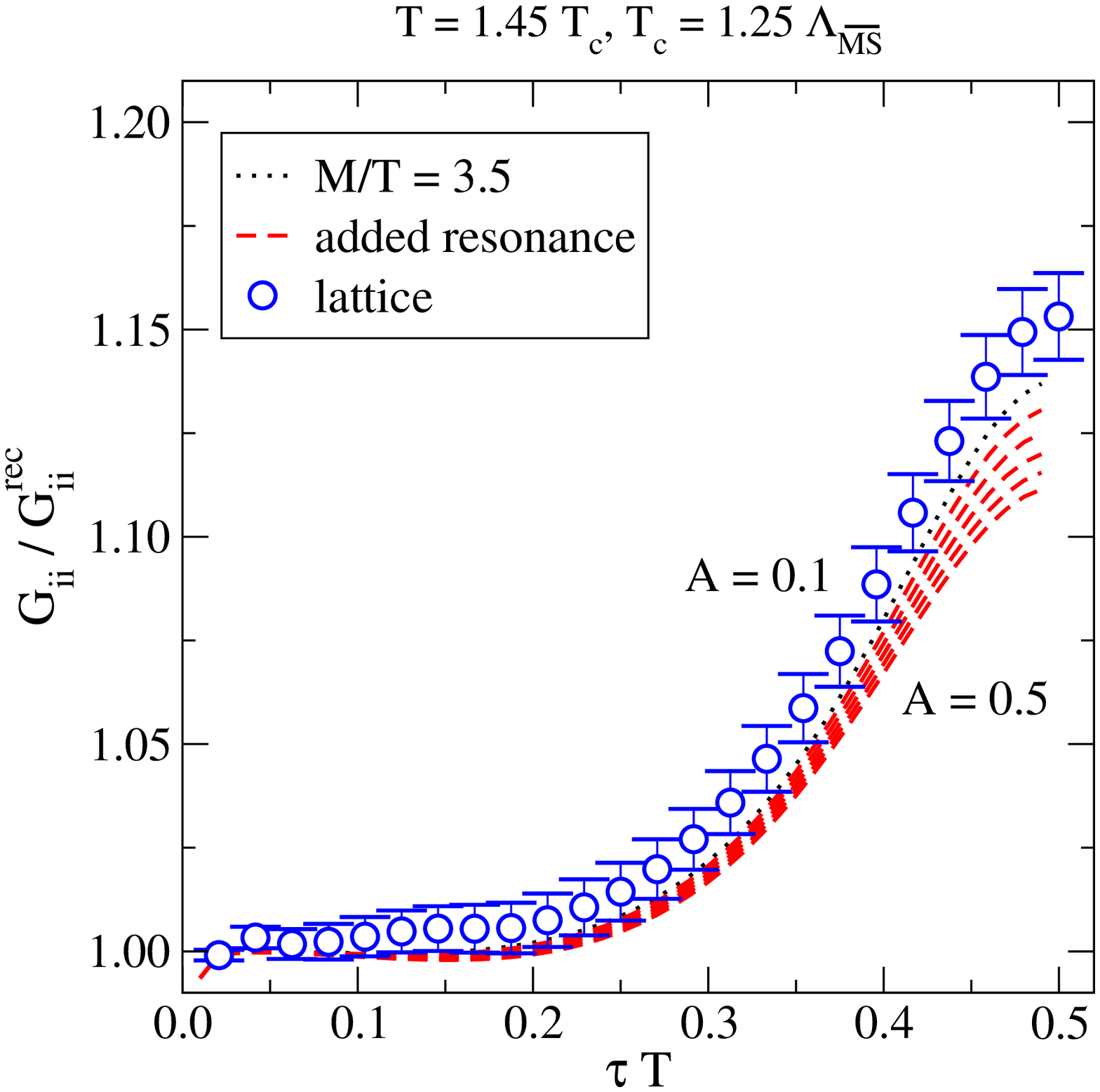}
~~~\epsfysize=7.5cm\epsfbox{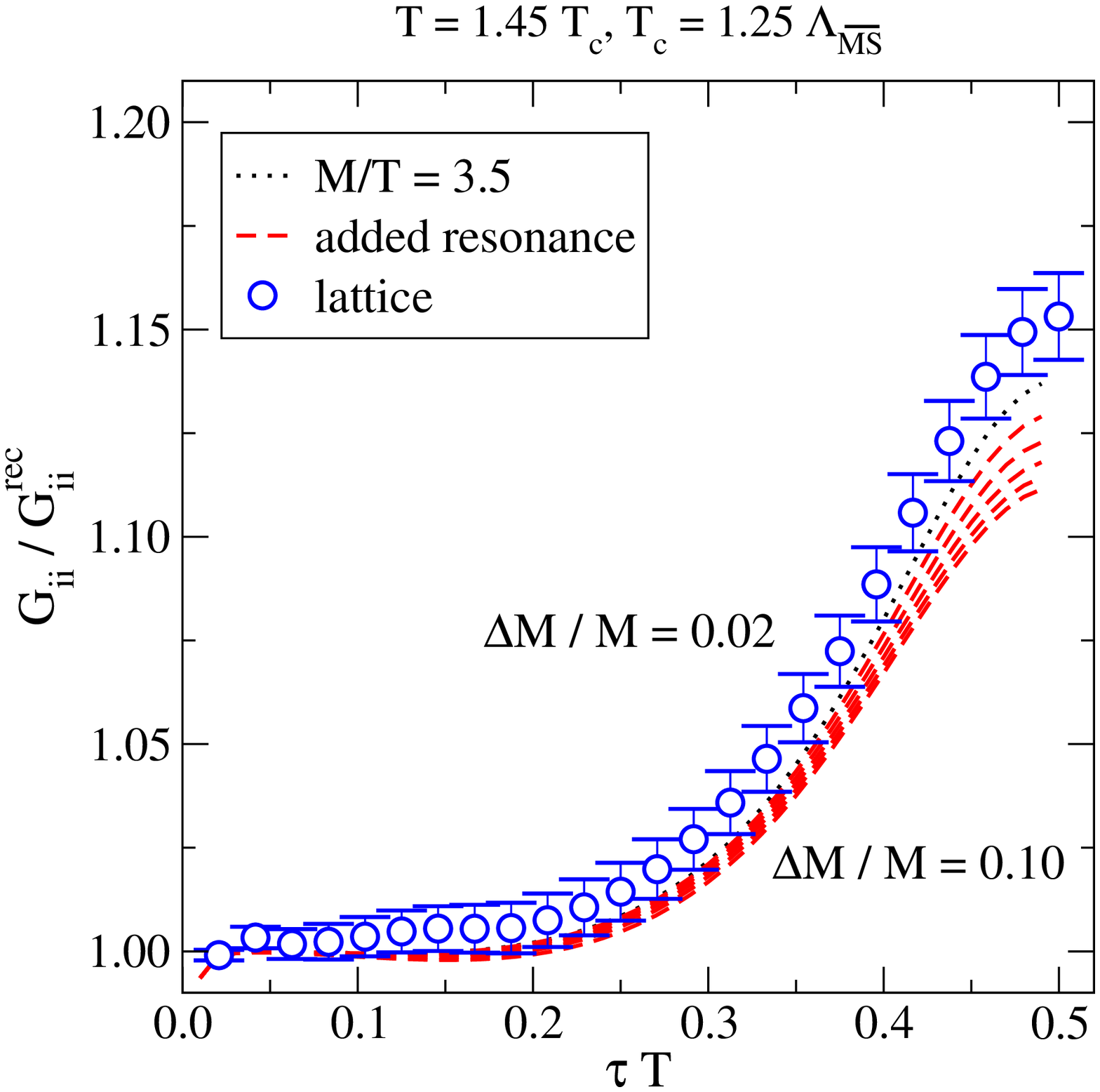}%
}

\caption[a]{\small
Left: The effect of a modified amplitude of a resonance peak, 
for $M/T = 3.5$, $\Delta M/M =0.1$, and 
$A = 0.1$, $0.2$, $0.3$, $0.4$, $0.5$ (cf.\ discussion below \eq\nr{BW}).
Right: The effect of a modified shape of a resonance peak, for $A = 0.5$ and
$\Delta M / M = 0.02,0.04,0.06,0.08,0.10$. 
}

\la{fig:threshold}
\end{figure*}

Based on our earlier investigations~\cite{pert,peskin},
we expect that in the temperature range of interest there is 
at most one resonance peak in the vector channel spectral function, 
placed slightly to the left from the free quark-antiquark threshold. 
Akin to \eq\nr{Lorentz}, we model this by 
a skewed Breit-Wigner shape,\footnote{%
 The approach in this section is schematic; for possible 
 other model shapes see, e.g., ref.~\cite{mshl}. (A general 
 discussion of the sum rule approach can be found in 
 ref.~\cite{sumrule}.)
 } 
\be
 \rho^\rmii{(BW)}_{ii}(\omega) \; \equiv \;  
 \frac{A \, \omega^2 \gamma^2}{(\omega')^2 + \gamma^2} 
 \frac{1}{ \cosh(\frac{\omega'}{2 M}) }
 \;, \quad \omega' \equiv  \omega - 2 M  + \Delta M 
 \;, \la{BW} 
\ee
constructed so as not to contribute to the transport coefficient. 
To reduce the number of free parameters to two, 
we set $\Delta M \equiv 2 \gamma$ in the following.
The contribution from such a peak, determined through \eq\nr{relation}, 
is added to the thermal NLO result as such, and to 
the vacuum result of \eq\nr{full_vac} with $A\to 5A$,  
$\gamma\to\gamma/5$, keeping the area under the peak roughly invariant. 
Obviously
these choices are arbitrary, but they should nevertheless convey 
a qualitative impression on the importance of resonance contributions. 
Based on ref.~\cite{nlo}, in which a resonance peak around a threshold
was matched to the thermal NLO spectral function above the threshold, 
we expect that $A \lsim 0.5$ and $\Delta M / M \lsim 0.1$.  

Results are shown in \fig\ref{fig:threshold}. 
Comparing with \fig\ref{fig:transport}, it can be seen that 
a change of the spectral shape around the threshold region 
affects the Euclidean correlator at smaller $\tau$ than 
a change of the transport peak. 
In fact, if the resolution, after continuum extrapolation, were 
high enough that the lattice and perturbative curves could be {\em subtracted}
from each other (rather than normalizing them to a function which diverges
at short distances), then the two features could be disentangled by 
inspecting intermediate distances, $0.2 < \tau < 0.3$, in which the 
threshold region contributes much more prominently than the transport peak.
In contrast the two features are difficult to tell apart if reliable 
data is only available for $0.3 < \tau < 0.5$.
(That said, we stress again 
that the present section is meant as indicative only.) 

%
\section{Conclusions}
\la{se:concl}

The main purpose of this paper has been to compute the massive
quark-number susceptibility and vector current correlator at 
next-to-leading order (NLO) in thermal QCD. 
The NLO results are shown in \eqs\nr{susc_final},
\nr{NLO_tau_final}, 
\nr{NLO_const_final}, and illustrated 
numerically in \figs\ref{fig:susc}, \ref{fig:Gii}. 

Our semi-analytic results can be directly compared with 
numerical lattice Monte Carlo simulations. Although no continuum
limit has been reached for the massive vector current 
correlator~\cite{ding2}, the 
agreement as seen in \figs\ref{fig:susc}, \ref{fig:Gii} 
is quite remarkable. 
(We have compared with quenched data, because these
are ``closest'' to the continuum limit, but our analytic 
results are also valid for the unquenched case.)

The good agreement suggests
that resummed perturbative NLO computations, through which the Minkowskian 
spectral function has been determined around the threshold
region~\cite{pert,peskin}, might be more accurate 
than sometimes assumed. A similar observation has also been made by 
comparing NLO computations to lattice data at spatial separations
relevant for quarkonium physics~\cite{singlet}.

The back side of a remarkable agreement is that the strict NLO 
result of the present paper reflects the physics of 
an infinitely narrow transport peak, rather than genuine 
heavy quark diffusion, and of a threshold singularity, rather than 
genuine quarkonium resonances --- and yet it works well. 
This implies that even though impressive 
efforts to extract properties of the transport peak
and quarkonium resonances from lattice data are being made, a very 
fine resolution is needed for getting 
systematic errors fully under control. 

In order to quantify the resolution that lattice simulations should
reach for observing substantial deviations from the NLO results, we have
finally probed  the significance of possible non-perturbative effects.
The impact of a smeared transport peak 
is illustrated in~\fig\ref{fig:transport}; the impact of 
resonance-like spectral weight below the threshold in \fig\ref{fig:threshold}. 
Interpreting lattice results such as those in 
ref.~\cite{ding2} as a deviation from NLO, 
it appears that a diffusion coefficient $2\pi T D \sim 1-2$ could be observed
in principle, but that larger values 
are hard to disentangle. 
Perhaps, a fruitful approach is 
to determine the diffusion coefficient 
by other means~\cite{Francis:2011gc,Banerjee:2011ra}, subtract 
the corresponding contribution from lattice data, and hope 
that any remaining deviations reflect the
physics of the threshold region. 

Ultimately, 
going beyond modelling, we suspect that
systematic studies are only possible once a continuum extrapolation
is available in a broad $\tau$-range. At this stage vacuum physics
can be subtracted, as outlined in ref.~\cite{cond}, 
and the non-divergent remainder subjected (at least in principle) 
to a model-independent analytic continuation~\cite{cuniberti}. 

%
\section*{Acknowledgements}

We thank H.-T.~Ding for helpful discussions and 
providing us with lattice data from ref.~\cite{ding2}, and
M.~Veps\"al\"ainen for collaboration at initial stages of this work. 
This work was partly supported by the Swiss National Science Foundation
(SNF) under grant 200021-140234.

%
\appendix
\renewcommand{\thesection}{Appendix~\Alph{section}}
\renewcommand{\thesubsection}{\Alph{section}.\arabic{subsection}}
\renewcommand{\theequation}{\Alph{section}.\arabic{equation}}

%
\section{Results for individual master sum-integrals}

In this appendix we list results for the 2-loop sum-integrals in 
\eqs\nr{nlo}, \nr{xnlo}, after carrying out the Matsubara sums as well 
as the Fourier transformation in \eq\nr{sum}. For brevity we refer to 
the sum-integrals with the notation  
\be
 \mathcal{I}^{m_1m_2m_3}_{n_1n_2n_3n_4n_5} (\tau) \equiv
 T \sum_{\omega_n} e^{-i \omega_n \tau}
 \left. 
 \Tint{K\{P\}}
  \frac{(M^2)^{m_1} (Q^2)^{m_2}(2K\cdot Q)^{m_3}}
 {(K^2)_{ }^{\raise0.4ex\hbox{$\scriptstyle n_1$}} 
 \Delta^{n_2}_P \Delta^{n_3}_{P-K}\Delta^{n_4}_{P-Q}
 \Delta^{n_5}_{P-K-Q}}
 \right|_{Q = (\omega_n,\vec{0})}
 \;. \la{def_I}
\ee
The index $m_1$ 
guarantees that all the masters have the same dimensionality. It 
is also convenient to introduce the shorthand notations
\be
 \Delta_{\sigma\tau} \equiv \epsilon_k + \sigma E_p + \tau E_{pk}
 \;, \quad
 \Delta_{\sigma} = E_p + \sigma E_{pk} 
 \;, \la{def_Delta}
\ee
where the energies are defined as in \eq\nr{energies}.
In some cases the remaining integrands can be simplified via 
the symmetrization $\vec{p}\leftrightarrow\vec{k-p}$. 

Four of the masters appearing
are independent of the external momentum $Q$, 
and therefore lead to contact terms: 
\ba
 \mathcal{I}^{000}_{12000}(\tau) & = & 
 \delta^{ }_\beta(\tau) \,
 \Tint{K\{P\}} \frac{1}{K^2 \Delta^2_P}
 \;, \la{e1} \\ 
 \mathcal{I}^{000}_{02100}(\tau) & = & 
 \delta^{ }_\beta(\tau) \,
 \Tint{K\{P\}} \frac{1}{\Delta^2_P \Delta^{ }_{P-K} }
 \;, \la{e2} \\ 
 \mathcal{I}^{000}_{11100}(\tau) & = & 
 \delta^{ }_\beta(\tau) \,
 \Tint{K\{P\}} \frac{1}{K^2 \Delta^{ }_P  \Delta^{ }_{P-K}}
 \;, \la{e3} \\ 
 \mathcal{I}^{100}_{12100}(\tau) & = & 
 \delta^{ }_\beta(\tau) \,
 \Tint{K\{P\}} \frac{M^2}{K^2 \Delta^2_P  \Delta^{ }_{P-K}}
 \;. \la{e4} 
\ea
No further details are given since these contact terms cancel  
against contributions from masters with $m_2 > 0$
(\eqs\nr{m7}, \nr{m11}, \nr{m14}, \nr{m16}, \nr{m22}).
After carrying out the Matsubara sums 
the remaining masters can be represented as  
(with the notation of \eq\nr{D_def})
\ba
 \mathcal{I}^{000}_{11010}(\tau) & = & 
 \int_{\vec{p,k}} 
 \frac{ 1+2 \nB{}(\epsilon_k) }{8 \epsilon_k E_p^2}
 \Bigl[ D^{ }_{2E_p}(\tau) + 2 T \nF{}'(E_p) \Bigr]
 \;, \la{m5} \\ 
 \mathcal{I}^{100}_{12010}(\tau) & = & 
 \int_{\vec{p,k}} \frac{M^2 [1+2 \nB{}(\epsilon_k)]
 }{32 \epsilon_k E_p^4} 
 \Bigl[ 
 2 D^{ }_{2E_p}(\tau) - E_p \partial_{E_p} D^{ }_{2E_p}(\tau) 
 \nn & & \qquad 
 +\, 4 T \nF{}'(E_p) - 2 T E_p \nF{}''(E_p)
 \Bigr]  
 \;, \la{m6} \\ 
 \mathcal{I}^{010}_{12010}(\tau) & = & 
 \mathcal{I}^{000}_{12000}(\tau) + 
 \int_{\vec{p,k}} 
 \frac{1+2 \nB{}(\epsilon_k)
 }{8 \epsilon_k E_p} 
 \Bigl[ 
   \partial_{E_p} D^{ }_{2E_p}(\tau) 
 \Bigr]  
 \;, \la{m7} \\ 
 \mathcal{I}^{000}_{10110}(\tau) & = & 
 \int_{\vec{p,k}} 
 \frac{1}{8 \epsilon_k E_p E_{pk}}
 \Bigl[
   D^{ }_{\epsilon_k E_p E_{pk}}(\tau) + 
   D^{\epsilon_k}_{E_p E_{pk}}(\tau) - 
   2 D^{E_p}_{\epsilon_k E_{pk}}(\tau) 
 \Bigr]
 \;, \la{m8} \\ 
 \mathcal{I}^{100}_{11110}(\tau) & = & 
 \int_{\vec{p,k}} 
 \frac{M^2}{8 \epsilon_k E_p E_{pk}}
 \biggl\{
   - \frac{ D^{ }_{\epsilon_k E_p E_{pk}}(\tau)}
   { \Delta^{ }_{++} \Delta^{ }_{-+} }  
   - \frac{ D^{\epsilon_k}_{E_p E_{pk}}(\tau) }
   { \Delta^{ }_{+-} \Delta^{ }_{--} } 
   + 
   \frac{ 2 \epsilon_k D^{E_p}_{\epsilon_k E_{pk}}(\tau)}
   {\Delta^{ }_{++} \Delta^{ }_{-+}\Delta^{ }_{--} } 
 \nn & & \qquad
   + \, \frac{  D^{ }_{2E_p}(\tau) + 2 T \nF{}'(E_p) }{E_p}  
   \biggl[
    \frac{\epsilon_k + E_{pk}}{\Delta^{ }_{++}\Delta^{ }_{-+} }
 \nn & & \qquad \qquad 
    - \,
   \epsilon_k\, \nF{}(E_{pk})
      \Bigl(
        \frac{1}{\Delta^{ }_{++}\Delta^{ }_{--}}
       + 
        \frac{1}{\Delta^{ }_{+-}\Delta^{ }_{-+}}
      \Bigr) 
 \nn & & \qquad \qquad 
      -\,
    E_{pk}\, \nB{}( \epsilon_k )
      \Bigl(
        \frac{1}{\Delta^{ }_{++}\Delta^{ }_{+-}}
       + 
        \frac{1}{\Delta^{ }_{-+}\Delta^{ }_{--}}
      \Bigr)  
   \biggr] 
 \biggr\}
 \;, \la{m9} \\ 
 \mathcal{I}^{001}_{11110}(\tau) & = & 
 \int_{\vec{p,k}} 
 \frac{1}{4 E_p E_{pk}}
 \biggl\{
    \frac{ D^{ }_{\epsilon_k E_p E_{pk}}(\tau)}
   { \Delta^{ }_{-+} }  
   + \frac{ D^{\epsilon_k}_{E_p E_{pk}}(\tau) }
   { \Delta^{ }_{+-} } 
   - 
   \frac{ 2 \epsilon_k D^{E_p}_{\epsilon_k E_{pk}}(\tau)}
   {\Delta^{ }_{++} \Delta^{ }_{--} } 
 \nn & & \qquad
    - \, 2 E_p  D^{ }_{2E_p}(\tau) \, 
   \biggl[
    \frac{1}{\Delta^{ }_{++}\Delta^{ }_{-+} }
 \nn & & \qquad \qquad 
    - \,
    \nF{}(E_{pk})
      \Bigl(
        \frac{1}{\Delta^{ }_{++}\Delta^{ }_{-+}}
       + 
        \frac{1}{\Delta^{ }_{+-}\Delta^{ }_{--}}
      \Bigr) 
 \nn & & \qquad \qquad 
      +\,
     \nB{}( \epsilon_k )
      \Bigl(
        \frac{1}{\Delta^{ }_{++}\Delta^{ }_{-+}}
       - 
        \frac{1}{\Delta^{ }_{+-}\Delta^{ }_{--}}
      \Bigr)  
   \biggr] 
 \biggr\}
 \;, \la{nm10} \\ 
 \mathcal{I}^{010}_{11110}(\tau) & = & 
 \mathcal{I}^{000}_{11100}(\tau) + 
 \int_{\vec{p,k}} 
 \frac{1}{8 \epsilon_k E_p E_{pk}}
 \biggl\{
    \frac{ \Delta^{ }_{++} }
   { \Delta^{ }_{-+} }\,  D^{ }_{\epsilon_k E_p E_{pk}}(\tau)
   + \frac{ \Delta^{ }_{--} }
   { \Delta^{ }_{+-} }\,  D^{\epsilon_k}_{E_p E_{pk}}(\tau)
 \nn & & \quad
   - \Bigl( \frac{\Delta^{ }_{-+}}{\Delta^{ }_{++}} + 
 \frac{\Delta^{ }_{-+}}{\Delta^{ }_{--}}
 \Bigr)
 \,  D^{E_p}_{\epsilon_k E_{pk}}(\tau)  
    - \, 4 E_p  D^{ }_{2E_p}(\tau) \, 
   \biggl[
    \frac{\epsilon_k + E_{pk}}{\Delta^{ }_{++}\Delta^{ }_{-+} }
 \nn & & \qquad \qquad 
    - \,
    \epsilon_k\, \nF{}(E_{pk})
      \Bigl(
        \frac{ 1 }{\Delta^{ }_{++}\Delta^{ }_{--}}
       + 
        \frac{ 1 }{\Delta^{ }_{+-}\Delta^{ }_{-+}}
      \Bigr) 
 \nn & & \qquad \qquad 
      -\, E_{pk} \, 
     \nB{}( \epsilon_k )
      \Bigl(
        \frac{ 1 }{\Delta^{ }_{++}\Delta^{ }_{+-}}
       + 
        \frac{ 1 }{\Delta^{ }_{-+}\Delta^{ }_{--}}
      \Bigr)  
   \biggr] 
 \biggr\}
 \;, \la{m11} \\ 
 \mathcal{I}^{000}_{01110}(\tau) & = & 
 \int_{\vec{p,k}} 
 \frac{ 1- 2 \nF{}( E_{pk} ) }{8 E_p^2 E_{pk}}
 \Bigl[ D^{ }_{2E_p}(\tau) + 2 T \nF{}'(E_p) \Bigr]
 \;, \la{m12} \\ 
 \mathcal{I}^{100}_{02110}(\tau) & = & 
 \int_{\vec{p,k}} 
  \frac{M^2 [1 - 2 \nF{}( E_{pk} )] }{32 E_p^4 E_{pk} } 
 \Bigl[ 
 2 D^{ }_{2E_p}(\tau) - E_p \partial_{E_p} D^{ }_{2E_p}(\tau) 
 \nn & & \qquad 
 +\, 4 T \nF{}'(E_p) - 2 T E_p \nF{}''(E_p)
 \Bigr]  
 \;, \la{m13} \\ 
 \mathcal{I}^{010}_{02110}(\tau) & = & 
 \mathcal{I}^{000}_{02100}(\tau) + 
 \int_{\vec{p,k}} 
 \frac{1 - 2 \nF{}( E_{pk} )
 }{8 E_p E_{pk} } 
 \Bigl[ 
   \partial_{E_p} D^{ }_{2E_p}(\tau) 
 \Bigr]  
 \;, \la{m14} \\ 
 \mathcal{I}^{200}_{12110}(\tau) & = & 
 \int_{\vec{p,k}} 
 \frac{ M^4 }{8 \epsilon_k E_p E_{pk}}
 \biggl\{
    \frac{ D^{ }_{\epsilon_k E_p E_{pk}}(\tau)  }
   { \Delta^{2}_{++} \Delta^{2}_{-+} }  
   + \frac{ D^{\epsilon_k}_{E_p E_{pk}}(\tau) }
   { \Delta^{2}_{+-}\Delta^{2}_{--} }  
   -   \frac{  D^{E_p}_{\epsilon_k E_{pk}}(\tau)  
 }{\Delta^{2}_{-+}}
 \Bigl(  
 \frac{ 1 }{\Delta^{2}_{++}}
  + 
 \frac{ 1 }{\Delta^{2}_{--}}
 \Bigr)
  \nn & & \qquad
    + \, \frac{ D^{ }_{2E_p}(\tau) + 2 T \nF{}'(E_p) }{2 E_p^3}  \, 
   \biggl[
   ( \epsilon_k + E_{pk} ) 
   \Bigl( 
     \frac{      \Delta^{ }_{++}\Delta^{ }_{-+}  
   - 2 E_p^2 }{\Delta^{2}_{++}\Delta^{2}_{-+} }
   \Bigr)
 \nn & & \qquad\qquad 
    - \,
    \epsilon_k\, \nF{}(E_{pk})
      \Bigl(
        \frac{\Delta^{ }_{++}\Delta^{ }_{--}
          - 2 E_p \Delta^{ }_+ }{\Delta^{2}_{++}\Delta^{2}_{--}}
       +  
        \frac{\Delta^{ }_{+-}\Delta^{ }_{-+} 
          - 2 E_p \Delta^{ }_- }{\Delta^{2}_{+-}\Delta^{2}_{-+}}
      \Bigr) 
 \nn & & \qquad  
      -\, E_{pk} \, 
     \nB{}( \epsilon_k )
      \Bigl(
        \frac{\Delta^{ }_{++}\Delta^{ }_{+-}
         +   2 E_p (\epsilon_k + E_p ) }{\Delta^{2}_{++}\Delta^{2}_{+-}}
       + 
        \frac{ \Delta^{ }_{-+}\Delta^{ }_{--}
         -  2 E_p (\epsilon_k - E_p) }{\Delta^{2}_{-+}\Delta^{2}_{--}}
      \Bigr)  
   \biggr] 
  \nn & & \qquad
    - \, \frac{  \partial^{ }_{E_p} 
   D^{ }_{2E_p}(\tau) + 2 T \nF{}''(E_p) }{4 E_p^2}  \, 
   \biggl[
     \frac{  \epsilon_k + E_{pk}  }{\Delta^{ }_{++}\Delta^{ }_{-+} } 
 \nn & & \qquad\qquad 
    - \,
    \epsilon_k\, \nF{}(E_{pk})
      \Bigl(
        \frac{1 }{\Delta^{ }_{++}\Delta^{ }_{--}}
       +  
        \frac{ 1 }{\Delta^{ }_{+-}\Delta^{ }_{-+}}
      \Bigr) 
 \nn & & \qquad\qquad  
      -\, E_{pk} \, 
     \nB{}( \epsilon_k )
      \Bigl(
        \frac{ 1 }{\Delta^{ }_{++}\Delta^{ }_{+-}}
       + 
        \frac{ 1 }{\Delta^{ }_{-+}\Delta^{ }_{--}}
      \Bigr)  
   \biggr] 
 \biggr\}
 \;, \la{m15} \\ 
 \mathcal{I}^{110}_{12110}(\tau) & = & 
 \mathcal{I}^{100}_{12100}(\tau) + 
 \int_{\vec{p,k}} 
 \frac{ M^2 }{8 \epsilon_k E_p E_{pk}}
 \biggl\{
   -  \frac{ D^{ }_{\epsilon_k E_p E_{pk}}(\tau)  }
   { \Delta^{2}_{-+} }  
   - \frac{ D^{\epsilon_k}_{E_p E_{pk}}(\tau) }
   { \Delta^{2}_{+-} }  
 \nn & & \qquad
   + \,  D^{E_p}_{\epsilon_k E_{pk}}(\tau)  
 \Bigl(  
 \frac{ 1 }{\Delta^{2}_{++}}
  + 
 \frac{ 1 }{\Delta^{2}_{--}}
 \Bigr)
    + \, 4 D^{ }_{2E_p}(\tau)  \, 
   \biggl[
     \frac{ 
    ( \epsilon_k + E_{pk} ) E_p  
   }{\Delta^{2}_{++}\Delta^{2}_{-+} }
  \nn & & \qquad \qquad
    - \,
    \epsilon_k\, \nF{}(E_{pk})
      \Bigl(
        \frac{ \Delta^{ }_+ }{\Delta^{2}_{++}\Delta^{2}_{--}}
       +  
        \frac{ \Delta^{ }_- }{\Delta^{2}_{+-}\Delta^{2}_{-+}}
      \Bigr) 
 \nn & & \qquad  \qquad
      +\, E_{pk}  \, 
     \nB{}( \epsilon_k )
      \Bigl(
        \frac{\epsilon_k + E_p  }{ \Delta^{2}_{++}\Delta^{2}_{+-}}
       - 
        \frac{ \epsilon_k - E_p }{\Delta^{2}_{-+}\Delta^{2}_{--}}
      \Bigr)  
   \biggr] 
  \nn & & \qquad
  +\,  \partial^{ }_{E_p} 
   D^{ }_{2E_p}(\tau)  \, 
   \biggl[
     \frac{  \epsilon_k + E_{pk}  }{\Delta^{ }_{++}\Delta^{ }_{-+} } 
 \nn & & \qquad  \qquad
    - \,
    \epsilon_k\, \nF{}(E_{pk})
      \Bigl(
        \frac{1 }{\Delta^{ }_{++}\Delta^{ }_{--}}
       +  
        \frac{ 1 }{\Delta^{ }_{+-}\Delta^{ }_{-+}}
      \Bigr) 
 \nn & & \qquad \qquad  
      -\, E_{pk} \, 
     \nB{}( \epsilon_k )
      \Bigl(
        \frac{ 1 }{\Delta^{ }_{++}\Delta^{ }_{+-}}
       + 
        \frac{ 1 }{\Delta^{ }_{-+}\Delta^{ }_{--}}
      \Bigr)  
   \biggr] 
 \biggr\}
 \;, \la{m16} \\ 
 \mathcal{I}^{100}_{01111}(\tau) & = & 
 \int_{\vec{p,k}} 
 \frac{ M^2 }{8 E_p^2 E_{pk}}
 \biggl\{ \frac{2T \nF{}'(E_p)\nF{}'(E_{pk}) }{ E_{pk} }
 \nn & & \qquad
 - \, \Bigl[ 1 - 2 \nF{}(E_{pk}) \Bigr]
 \biggl[  \frac{D^{ }_{2E_p}(\tau)}{\Delta^{ }_+\Delta^{ }_-} - 
   \frac{2 T \nF{}'(E_{p})}{E_{pk}^2}
 \biggr]
 \biggr\}
 \;, \la{m17} \\ 
 \mathcal{I}^{010}_{01111}(\tau) & = & 
 \int_{\vec{p,k}} 
 \frac{ 1 }{2 E_{pk}}
 \, \Bigl[ 1 - 2 \nF{}(E_{pk}) \Bigr]
 \frac{D^{ }_{2E_p}(\tau)}{\Delta^{ }_+\Delta^{ }_-}
 \;, \la{m18} \\ 
 \mathcal{I}^{000}_{-11111}(\tau) & = & 
 \int_{\vec{p,k}} 
 \frac{ 1 }{8 E_p^2 E_{pk}}
 \biggl\{
 \Bigl( \epsilon_k^2 - E_p^2 - E_{pk}^2 \Bigr) 
 \frac{2T \nF{}'(E_p)\nF{}'(E_{pk}) }{ E_{pk} }
 \nn & & \qquad
 - \, \Bigl[ 1 - 2 \nF{}(E_{pk}) \Bigr]
 \Bigl( \epsilon_k^2 - E_p^2 + E_{pk}^2 \Bigr) 
 \biggl[  \frac{D^{ }_{2E_p}(\tau)}{\Delta^{ }_+\Delta^{ }_-} - 
   \frac{2 T \nF{}'(E_{p})}{E_{pk}^2}
 \biggr]
 \biggr\}
 \;, \la{m19} \\ 
 \mathcal{I}^{200}_{11111}(\tau) & = & 
 \int_{\vec{p,k}} 
 \frac{ M^4 }{4 \epsilon_k E_p E_{pk}}
 \biggl\{
 \Bigl( \epsilon_k^2 - E_p^2 - E_{pk}^2 \Bigr) 
 \frac{\epsilon_k}{ E_p E_{pk} }
 \frac{ T \nF{}'(E_p)\nF{}'(E_{pk}) }
 {  \Delta^{ }_{++} \Delta^{ }_{+-} \Delta^{ }_{-+} \Delta^{ }_{--} }
 \nn & &  \qquad  + \, \frac{
   D^{ }_{\epsilon_k E_p E_{pk}}(\tau)
  }{ \Delta^{2}_{++} \Delta^{ }_{+-} \Delta^{ }_{-+} } 
  +  
   \frac{ 
   D^{\epsilon_k}_{E_p E_{pk}}(\tau)
  }{ \Delta^{2}_{--} \Delta^{ }_{+-} \Delta^{ }_{-+} } 
   -  \frac{2 
     D^{E_p}_{\epsilon_k E_{pk}}(\tau)  
  }{ \Delta^{2}_{-+}\Delta^{ }_{++} \Delta^{ }_{--} }  
  \nn & & \qquad
    - \, \frac{D^{ }_{2E_p}(\tau)}{2 E_p^2}  \, 
   \biggl[
     \frac{ 
     E_p (\epsilon_k + 2 E_{pk})    
   }{\Delta^{ }_+ \Delta^{ }_- \Delta^{ }_{++}\Delta^{ }_{-+} }
  \nn & & \qquad \qquad
    - \,
    \epsilon_k\, \nF{}(E_{pk})
      \Bigl(
        \frac{ 1 }{ \Delta^{ }_+\Delta^{ }_{++}\Delta^{ }_{--}}
       +  
        \frac{ 1 }{  \Delta^{ }_- \Delta^{ }_{+-}\Delta^{ }_{-+}}
      \Bigr) 
 \nn & & \qquad  \qquad
      +\, \frac{ E_{pk}  \, 
     \nB{}( \epsilon_k ) }{\epsilon_k}
      \Bigl(
        \frac{ 1 }{  \Delta^{ }_{++}\Delta^{ }_{+-}}
       - 
        \frac{ 1 }{  \Delta^{ }_{-+}\Delta^{ }_{--}}
      \Bigr)  
   \biggr]  
  \nn & & \qquad
    + \, \frac{2 T \nF{}'(E_p)}{E_p}  \, 
   \biggl[
     \frac{1}{4 E_{pk}^2}
    \Bigl(
       \frac{\Delta^{ }_{++} + E_{pk}}{\Delta^{2 }_{++}}  + 
       \frac{\Delta^{ }_{-+} + E_{pk}}{\Delta^{2 }_{-+}}
    \Bigr)
  \nn & & \qquad \qquad
    - \,
    \frac{ \epsilon_k\, \nF{}(E_{pk}) }{2 E_{pk}^2}
      \Bigl(
        \frac{ \epsilon_k^2 + E_p^2 - E_{pk}^2 - 2 \Delta_+^2 }
   {  \Delta^{2}_{++}\Delta^{2}_{--}}
       +  
        \frac{ \epsilon_k^2 + E_p^2 - E_{pk}^2 - 2 \Delta_-^2  }
  {  \Delta^{2}_{+-}\Delta^{2}_{-+}}
      \Bigr) 
 \nn & & \qquad  \qquad
      +\, E_{pk}  \, 
     \nB{}( \epsilon_k )
      \Bigl(
        \frac{ 1 }{\Delta^{2}_{++}\Delta^{2}_{+-}}
       + 
        \frac{ 1 }{\Delta^{2}_{-+}\Delta^{2}_{--}}
      \Bigr)  
   \biggr]  
 \biggr\}
 \;, \la{m20} \\ 
 \mathcal{I}^{110}_{11111}(\tau) & = & 
 \int_{\vec{p,k}} 
 \frac{ M^2 }{4 \epsilon_k E_p E_{pk}}
 \biggl\{
  - \frac{
   D^{ }_{\epsilon_k E_p E_{pk}}(\tau)
  }{ \Delta^{ }_{+-} \Delta^{ }_{-+} } 
  -  
   \frac{ 
   D^{\epsilon_k}_{E_p E_{pk}}(\tau)
  }{ \Delta^{ }_{+-} \Delta^{ }_{-+} } 
   + \, \frac{2 
     D^{E_p}_{\epsilon_k E_{pk}}(\tau)  
  }{ \Delta^{ }_{++} \Delta^{ }_{--} }  
  \nn & & \qquad
    + \, 2 D^{ }_{2E_p}(\tau)  \, 
   \biggl[
     \frac{ 
     E_p (\epsilon_k + 2 E_{pk})
   }{\Delta^{ }_+ \Delta^{ }_- \Delta^{ }_{++}\Delta^{ }_{-+} }
  \nn & & \qquad \qquad
    - \,
    \epsilon_k\, \nF{}(E_{pk})
      \Bigl(
        \frac{ 1 }{ \Delta^{ }_+\Delta^{ }_{++}\Delta^{ }_{--}}
       +  
        \frac{ 1 }{  \Delta^{ }_- \Delta^{ }_{+-}\Delta^{ }_{-+}}
      \Bigr) 
 \nn & & \qquad  \qquad
      +\, \frac{ E_{pk}  \, 
     \nB{}( \epsilon_k ) }{ \epsilon_k } 
      \Bigl(
        \frac{ 1 }{  \Delta^{ }_{++}\Delta^{ }_{+-}}
       - 
        \frac{ 1 }{  \Delta^{ }_{-+}\Delta^{ }_{--}}
      \Bigr)  
   \biggr]  
 \biggr\}
 \;, \la{m21} \\ 
 \mathcal{I}^{020}_{11111}(\tau) & = & 
 2 \, \mathcal{I}^{000}_{11100}(\tau)
 + 
 \int_{\vec{p,k}} 
 \frac{ 1 }{4 \epsilon_k E_p E_{pk}}
 \biggl\{
   \frac{ \Delta^2_{++} 
   D^{ }_{\epsilon_k E_p E_{pk}}(\tau)
  }{ \Delta^{ }_{+-} \Delta^{ }_{-+} } 
  + 
   \frac{ \Delta^2_{--}
   D^{\epsilon_k}_{E_p E_{pk}}(\tau)
  }{ \Delta^{ }_{+-} \Delta^{ }_{-+} } 
 \nn & & \qquad
   - \, \frac{2 \Delta^2_{-+}
     D^{E_p}_{\epsilon_k E_{pk}}(\tau)  
  }{ \Delta^{ }_{++} \Delta^{ }_{--} }  
    - \, 8 E_p^2 D^{ }_{2E_p}(\tau)  \, 
   \biggl[
     \frac{ 
     E_p  (\epsilon_k + 2 E_{pk})
   }{\Delta^{ }_+ \Delta^{ }_- \Delta^{ }_{++}\Delta^{ }_{-+} }
  \nn & & \qquad \qquad
    - \,
    \epsilon_k\, \nF{}(E_{pk})
      \Bigl(
        \frac{ 1 }{ \Delta^{ }_+\Delta^{ }_{++}\Delta^{ }_{--}}
       +  
        \frac{ 1 }{  \Delta^{ }_- \Delta^{ }_{+-}\Delta^{ }_{-+}}
      \Bigr) 
 \nn & & \qquad  \qquad
      +\, \frac{ E_{pk}  \, 
     \nB{}( \epsilon_k ) }{ \epsilon_k }
      \Bigl(
        \frac{ 1 }{ \Delta^{ }_{++}\Delta^{ }_{+-}}
       - 
        \frac{ 1 }{  \Delta^{ }_{-+}\Delta^{ }_{--}}
      \Bigr)  
   \biggr]  
 \biggr\}
 \;. \la{m22}
\ea

%
\section{Renormalization of the complete result}

Inserting the expressions for the master sum-integrals from 
appendix~A into \eqs\nr{ct}, \nr{nlo}, \nr{xct}, \nr{xnlo},
we obtain renormalized results for the correlators considered.  
Like in \se\ref{se:qual}, the result can be divided into 
$\tau$-dependent and constant terms. The former reads
\ba
 & & \hspace*{-1cm}
 \frac{ 
  \left. G_\rmii{V}^\rmii{NLO} \right|_\rmii{$\tau$-dep.}
  }{ 
  4 g^2 C_A C_F 
  } = 
 \nn
 & = &
 \int_{\vec{p,k}} 
 \frac{D^{ }_{\epsilon_k E_p E_{pk}}(\tau)}{\epsilon_k E_p E_{pk}
 \Delta^{ }_{+-}\Delta^{ }_{-+} }
 \biggl\{ - E_p^2 - E_{pk}^2
  + M^2 \biggl[ 
  \frac{\Delta^{ }_{--}}{\Delta^{ }_{++}}
 + \frac{2\epsilon_k^2}{\Delta^{ }_{+-}\Delta^{ }_{-+}}
 \biggr]
 + \frac{4\epsilon_k^2 M^4}{\Delta_{++}^2 \Delta^{ }_{+-}\Delta^{ }_{-+}}
 \biggr\}
 \nn 
 & + &
 \int_{\vec{p,k}} 
 \frac{D^{\epsilon_k}_{E_p E_{pk}}(\tau)}{\epsilon_k E_p E_{pk}
 \Delta^{ }_{+-}\Delta^{ }_{-+}
 }
 \biggl\{ - E_p^2 - E_{pk}^2
  + M^2 \biggl[ 
  \frac{\Delta^{ }_{++}}{\Delta^{ }_{--}}
 + \frac{2\epsilon_k^2}{\Delta^{ }_{+-}\Delta^{ }_{-+}}
 \biggr]
 + \frac{4\epsilon_k^2 M^4}{\Delta_{--}^2 \Delta^{ }_{+-}\Delta^{ }_{-+}}
 \biggr\}
 \nn 
 & + &
 \int_{\vec{p,k}} 
 \frac{2 D^{E_p}_{\epsilon_k E_{pk}}(\tau)}{\epsilon_k E_p E_{pk}
  \Delta^{ }_{++}\Delta^{ }_{--}
 }
 \biggl\{ + E_p^2 + E_{pk}^2
  - M^2 \biggl[ 
  \frac{\Delta^{ }_{+-}}{\Delta^{ }_{-+}}
 + \frac{2\epsilon_k^2}{\Delta^{ }_{++}\Delta^{ }_{--}}
 \biggr]
 - \frac{4\epsilon_k^2 M^4}{\Delta_{-+}^2 \Delta^{ }_{++}\Delta^{ }_{--}}
 \biggr\}
 \nn 
 & + & \int_\vec{p} D^{ }_{2E_p}(\tau) \biggl\{ \mbox{``\eq\nr{coeff}''}
 \nn 
 &  &  
 \; + \, 
 \int_\vec{k} \frac{\nB{}(\epsilon_k)}{\epsilon_k} \biggl[ 
  -\frac{1}{E_p^2} + \frac{M^2}{E_p^4}
  + \frac{M^2(2 E_p^2 + M^2)}{2E_p^3E^{ }_{pk}}
  \biggl( \frac{1}{\Delta_{++}^2} 
        + \frac{1}{\Delta_{--}^2}
        - \frac{1}{\Delta_{+-}^2} 
        - \frac{1}{\Delta_{-+}^2}  \biggr)
 \nn 
 &  & \qquad\qquad - \,  
 \frac{2 (E_p^2 + E_{pk}^2 - M^2)}{E_p E^{ }_{pk}}
 \biggl( 
 \frac{1}{\Delta^{}_{++}\Delta^{ }_{--}} 
 - 
 \frac{1}{\Delta^{}_{+-}\Delta^{ }_{-+}} 
 \biggr)
 \nn 
 &  & \qquad\qquad + \,  
 \frac{M^2 (2 E_p^2 - M^2)}{E_p^4}
 \biggl( 
 \frac{1}{\Delta^{}_{++}\Delta^{ }_{--}}
 + 
 \frac{1}{\Delta^{}_{+-}\Delta^{ }_{-+}}
 \biggr)
 \biggr]
 \nn &  & \; + \,  
 \int_\vec{k} \frac{\nF{}(E_{pk})}{E_{pk}} \, \mathbbm{P}\biggl[ 
  \frac{1}{2E_p^2} + \frac{M^2}{E_p^4}
  + \frac{M^2(2 E_p^2 + M^2)}{2E_p^3\epsilon^{ }_k}
  \biggl( \frac{1}{\Delta_{-+}^2} 
        + \frac{1}{\Delta_{--}^2}
        - \frac{1}{\Delta_{+-}^2}
        - \frac{1}{\Delta_{++}^2} 
  \biggr)
 \nn &   & 
 \qquad\qquad 
  - \, 
 \frac{2(E_p^2 + E_{pk}^2 - M^2)}{ E_p} 
 \biggl(
   \frac{1}{\Delta^{ }_+\Delta^{ }_{++}\Delta^{ }_{--}}
   + \frac{1}{\Delta^{ }_-\Delta^{ }_{+-}\Delta^{ }_{-+}}
 \biggr)
 +  \frac{4 (E_p^2 + E_{pk}^2)-\epsilon_k^2}
         {2 E_p^2\, \Delta^{ }_+ \Delta^{ }_-} 
 \nn &   & \qquad\qquad + \,  
 \frac{E_{pk} M^2 (2 E_p^2  - M^2)}{E_p^4} 
 \biggl(
   \frac{1}{\Delta^{ }_+\Delta^{ }_{++}\Delta^{ }_{--}}
   - \frac{1}{\Delta^{ }_-\Delta^{ }_{+-}\Delta^{ }_{-+}}
 \biggr)
 \biggr]
 \biggr\} 
 \nn 
 & + &  \int_\vec{p} E_p \partial^{ }_{E_p} D^{ }_{2E_p}(\tau) 
 \biggl\{ 0 
 \nn & & \; + \, 
 \int_\vec{k} \frac{\nB{}(\epsilon_k)}{\epsilon_k} \biggl[ 
 -\frac{1}{E_p^2} - \frac{M^2}{2 E_p^4}
 + 
 \frac{M^2(2 E_p^2 + M^2)}
 {2E_p^4}
 \biggl( 
   \frac{1}{ \Delta^{ }_{++}\Delta^{ }_{+-} }
 +
   \frac{1}{ \Delta^{ }_{--}\Delta^{ }_{-+}}
 \biggr)
 \biggr]
 \nn &  & \; + \,  
 \int_\vec{k} \frac{\nF{}(E_{pk})}{E_{pk}} \biggl[ 
 -\frac{1}{E_p^2} - \frac{M^2}{2 E_p^4}
 + 
 \frac{M^2(2 E_p^2 + M^2)}{2E_p^4}
 \biggl( 
   \frac{1}{ \Delta^{ }_{++}\Delta^{ }_{--} }
 +
   \frac{1}{ \Delta^{ }_{+-}\Delta^{ }_{-+}}
 \biggr)
 \biggr]
 \biggr\} 
 \;. \la{NLO_tau}
\ea

The constant contribution, in turn, can be expressed as 
\ba
 & & \hspace*{-1cm}
 \frac{ 
  \left. G_\rmii{V}^\rmii{LO} \right|_\rmii{const.}
  }{ 
  4 g^2 C_A C_F 
  } =
 \nonumber \\[2mm]
 & = &
 \int_{\vec{p,k}} 
 \frac{T \nF{}'(E_p)\nF{}'(E_{pk})}{2 E_p^2 E_{pk}^2}
 \biggl\{
     -\epsilon_k^2 + E_p^2 + E_{pk}^2 
     + 2M^4 \biggl( 
   \frac{1}{ \Delta^{ }_{++}\Delta^{ }_{--} }
 +
   \frac{1}{ \Delta^{ }_{+-}\Delta^{ }_{-+}}
 \biggr)
 \biggr\}
 \nn 
 & + & \int_\vec{p} 2 T \nF{}'(E_p) \biggl\{ \mbox{``\eq\nr{coeff2}''}
 \nn & & \; + \, 
 \int_\vec{k} \frac{\nB{}(\epsilon_k)}{\epsilon_k} \biggl[ 
  -\frac{1}{E_p^2} + \frac{M^2}{E_p^4}
 - 
 \frac{2 \epsilon_k M^4}{E_p^3 }
 \biggl(
     \frac{1}{\Delta_{++}^2\Delta_{+-}^2}
  - 
     \frac{1}{\Delta_{--}^2\Delta_{-+}^2}
 \biggr)
 \nn 
 &  & \qquad\qquad + \, 
 \frac{M^2(2E_p^2 - M^2)}{E_p^4 }
 \biggl(
    \frac{1}{\Delta^{ }_{++}\Delta^{ }_{+-}} 
  +
    \frac{1}{\Delta^{ }_{--}\Delta^{ }_{-+}} 
 \biggr)
 \biggr]
 \nn &  & \; + \, 
 \int_\vec{k} \frac{\nF{}(E_{pk})}{E_{pk}} \biggl[ 
  - \frac{1}{E_p^2} + \frac{M^2}{E_p^4}
  + \frac{\epsilon_k^2 - E_p^2 - E_{pk}^2}{2 E_p^2 E_{pk}^2}
 + 
  \frac{2M^4}{ E_p^3 E^{ }_{pk} }
 \biggl(
    \frac{\Delta_+^2}{\Delta_{++}^2\Delta_{--}^2}
  -  \frac{\Delta_-^2}{\Delta_{+-}^2\Delta_{-+}^2}
 \biggr)
 \nn 
 &  & \qquad\qquad + \, 
 \biggl( 
 \frac{M^2 ( 2 E_p^2 -M^2)} {E_p^4} 
  - \frac{M^4} {E_p^2 E_{pk}^2} 
 \biggr)
 \biggl( \frac{1}{\Delta^{ }_{++} \Delta^{ }_{--}}
 + \frac{1}{\Delta^{ }_{+-} \Delta^{ }_{-+}} \biggr)
 \biggr]
 \biggr\} 
 \nn 
 & + & \int_\vec{p} 2 T E_p \nF{}''(E_p) \biggl\{ 0 
 \nn & & \; + \, 
 \int_\vec{k} \frac{\nB{}(\epsilon_k)}{\epsilon_k} \biggl[ 
 -\frac{M^2}{2 E_p^4} 
  +   
 \frac{M^4} {2 E_p^4}
 \biggl( 
   \frac{1}{ \Delta^{ }_{++}\Delta^{ }_{+-} }
 +
   \frac{1}{ \Delta^{ }_{--}\Delta^{ }_{-+}}
 \biggr)
 \biggr]
 \nn &  & \; + \,  
 \int_\vec{k} \frac{\nF{}(E_{pk})}{E_{pk}} \biggl[ 
 -\frac{M^2}{2 E_p^4} 
  +      
 \frac{M^4}{2 E_p^4} 
 \biggl( 
   \frac{1}{ \Delta^{ }_{++}\Delta^{ }_{--} }
 +
   \frac{1}{ \Delta^{ }_{+-}\Delta^{ }_{-+}}
 \biggr)
 \biggr]
 \biggr\} 
 \;. \la{NLO_const}
\ea
Finally, for the susceptibility, 
the NLO term amounts to
\ba
 \frac{ 
   G_{00}^\rmii{NLO} 
  }{ 
  4 g^2 C_A C_F 
  }
 & = &
 \int_{\vec{p,k}} 
 \frac{T \nF{}'(E_p)\nF{}'(E_{pk})}{E_p^{ }E_{pk}^{ }}
 \biggl\{
     -M^2 
 \biggl( 
   \frac{1}{ \Delta^{ }_{++}\Delta^{ }_{--} }
 -
   \frac{1}{ \Delta^{ }_{+-}\Delta^{ }_{-+}}
 \biggr)
 \biggr\}
 \nn 
 & + & \int_\vec{p} 2 T E_p \nF{}''(E_p) \biggl\{ 0 
 \nn & & \; + \, 
 \int_\vec{k} \frac{\nB{}(\epsilon_k)}{\epsilon_k} \biggl[ 
 -\frac{1}{2 E_p^2} 
  +   
 \frac{M^2} {2 E_p^2}
 \biggl( 
   \frac{1}{ \Delta^{ }_{++}\Delta^{ }_{+-} }
 +
   \frac{1}{ \Delta^{ }_{--}\Delta^{ }_{-+}}
 \biggr)
 \biggr]
 \nn &  & \; + \,  
 \int_\vec{k} \frac{\nF{}(E_{pk})}{E_{pk}} \biggl[ 
 -\frac{1}{2 E_p^2} 
  +   
 \frac{M^2}{2 E_p^2} 
 \biggl( 
   \frac{1}{ \Delta^{ }_{++}\Delta^{ }_{--} }
 +
   \frac{1}{ \Delta^{ }_{+-}\Delta^{ }_{-+}}
 \biggr)
 \biggr]
 \biggr\} 
 \;. \la{xNLO_const}
\ea

The ``0''s in \eqs\nr{NLO_tau}, \nr{NLO_const}, \nr{xNLO_const} 
represent vacuum contributions that vanish after renormalization. 
The coefficients of $\partial_{E_p}^{ }D_{2E_p}^{ }(\tau)$
and $T\nF{}''(E_p)$ 
vanish already when \eqs\nr{ct}, \nr{nlo} are summed together, but
the coefficients of $D_{2E_p}(\tau)$ and $T\nF{}'(E_p)$ not. 
They can be expressed as
(a principal value prescription is implied where necessary) 
\ba
 \mbox{``\eq\nr{coeff}''} & = & \int_\vec{k} \biggl\{ 
  \frac{ 1-\epsilon }{2 E_p^2 E^{ }_{pk}}
 -\frac{(4 - 4\epsilon + \frac{M^2}{E_p^2} )(\epsilon_k + E_{pk})} 
    {\epsilon_k E_{pk}[(\epsilon_k + E_{pk})^2 - E_p^2]}
 +  \frac{ 2 (1-\epsilon)^2 }{E^{ }_{pk}
           [(\epsilon_k + E_{pk})^2 - E_p^2]}
 \nn & & -\, 
  \frac{2 (2 - 2\epsilon + \frac{M^2}{E_p^2})  (\epsilon_k + E_{pk}) M^2}
         {\epsilon_k E^{ }_{pk}
         [(\epsilon_k + E_{pk})^2 - E_p^2]^2} 
 + 
  \frac{
     2 (1-\epsilon)(2+\epsilon)  + \frac{ \epsilon {M^2} }{ E_p^2}   
  }{2 E^{ }_{pk}
        (E_{pk}^2 - E_p^2)}
 \nn & & 
   - \,  \frac{(1-\epsilon)(\epsilon_k^2 + E_{pk}^2 - E_p^2)}
  {4 E_p^2 E^{ }_{pk}
        (E_{pk}^2 - E_p^2)}
   - 
       \frac{
       (
       4 - 4\epsilon + \frac{2\epsilon M^2}{E_p^2} - \frac{M^4}{E_p^4} 
       )
       (\epsilon_k + 2 E^{ }_{pk}) E_p^2}
        {\epsilon_k E^{ }_{pk}
         [(\epsilon_k + E_{pk})^2 - E_p^2](E_{pk}^2 - E_p^2)}
 \biggr\} \;, \hspace*{1cm} \la{coeff} \\
 \mbox{``\eq\nr{coeff2}''} & = & \int_\vec{k} \biggl\{ 
   \frac{1-\epsilon}{2 E_p^2 E^{ }_{pk}}
   -\frac{(\epsilon_k + E_{pk})M^2} 
    {\epsilon_k E_p^2 E^{ }_{pk}[(\epsilon_k + E_{pk})^2 - E_p^2]} 
   \nn & & 
   -\,   \frac{2  (\epsilon_k + E_{pk}) M^4}
         {\epsilon_k E_p^2 E_{pk}^{ }
         [(\epsilon_k + E_{pk})^2 - E_p^2]^2} 
   + \frac{\epsilon M^2}{2 E_p^2 E_{pk}^{3}}
   \nn & & 
   -\, \frac{(1-\epsilon)(\epsilon_k^2 + E_{pk}^2 - E_p^2)}
   { 4 E_p^2 E^{3}_{pk}}
   + \frac{[(\epsilon_k + 2 E^{ }_{pk})(\epsilon_k + E_{pk}^{ })^2
     - \epsilon_k E_p^2]
    M^4}{\epsilon_k E_p^2 E_{pk}^3 
    [(\epsilon_k + E_{pk})^2 - E_p^2]^2} 
 \biggr\} \;. \hspace*{1cm} \la{coeff2} 
\ea
The various structures here 
can be identified as specific vacuum integrals: 
\ba
 \int_\vec{k} \frac{1}{2\epsilon_k} 
 & = & 
 \int_{K} \frac{1}{K^2}
 \;, \la{h1} \\ 
 \int_\vec{k} \frac{1}{2E_{pk}} 
 & = & 
 \int_{K} \frac{1}{\Delta^{ }_{P-K}}
 \;, \la{h2} \\ 
 \int_\vec{k} \frac{1}{2 \epsilon_k E_{pk}} 
 \frac{\epsilon_k + E_{pk}}{(\epsilon_k + E_{pk})^2 - E_p^2}
 & = &
 \biggl[ \int_{K} \frac{1}{K^2 \Delta^{ }_{P-K}} \biggr]_{p_0 = i E_p}
 \;, \la{h3} \\ 
  \int_\vec{k} \frac{-E_p^2}{2 E_{pk}
           [(\epsilon_k + E_{pk})^2 - E_p^2]}
  & = & 
 \biggl[ p_0  \int_{K} \frac{k_0}{K^2 \Delta^{ }_{P-K}} \biggr]_{p_0 = i E_p}
 \;, \la{h4} \\ 
   \int_\vec{k} \frac{-E_p^2 (\epsilon_k + E_{pk})}
         {2 \epsilon_k E_{pk}
         [(\epsilon_k + E_{pk})^2 - E_p^2]^2}
  & = &
  \biggl[ p_0  \int_{K}
        \frac{p_0 - k_0}{K^2 \Delta^2_{P-K}} \biggr]_{p_0 = i E_p}
 \;, \la{h5} \\ 
  \int_\vec{k} \frac{1}{4 E_{pk}^3} 
  & = &
  \biggl[ \int_{K} \frac{1}{\Delta^2_{P-K}} \biggr]_{p_0 = i E_p}
 \;, \la{h7} \\ 
  \int_\vec{k} \frac{1}{4 E_{pk}
        (E_{pk}^2 - E_p^2)}
  & = &
  \biggl[ \int_{K}
        \frac{1}{\Delta^{ }_{P-K}\Delta^{ }_{P-K-Q}} \biggr]_
              {Q = (2 i E_p,\vec{0})}
 \;, \la{h6} \\ 
  \int_\vec{k} \frac{\epsilon_k^2 + E_{pk}^2 - E_p^2}{4 E_{pk}^3} 
  & = &
  \biggl[ \int_{K} \frac{K^2}{\Delta^2_{P-K}} \biggr]_{p_0 = i E_p}
 \la{h9} \;, \\ 
 \int_\vec{k} \frac{\epsilon_k^2 + E_{pk}^2 - E_p^2}{4 E_{pk}
        (E_{pk}^2 - E_p^2)}
  & = &
  \biggl[ \int_{K}
        \frac{K^2}{\Delta^{ }_{P-K}\Delta^{ }_{P-K-Q}} \biggr]_
              {p_0 = i E_p,Q = (2 i E_p,\vec{0})}
 \;, \hspace*{1cm} \la{h8} \\ 
  \int_\vec{k} \frac{(\epsilon_k + 2 E_{pk})(\epsilon_k + E_{pk})^2- 
          \epsilon_k E_p^2}
         {4 \epsilon_k E_{pk}^3
         [(\epsilon_k + E_{pk})^2 - E_p^2]^2}
  & = &
  \biggl[ \int_{K}
        \frac{1}{K^2\Delta^2_{P-K}} \biggr]_
              {p_0 = i E_p}
 \la{h10} \;, \\ 
  \int_\vec{k} \frac{\epsilon_k + 2 E_{pk}}
   {4 \epsilon_k E_{pk}[(\epsilon_k + E_{pk})^2 - E_p^2](E_{pk}^2 - E_p^2)}
  & = & \biggl[ \int_{K}
        \frac{1}{K^2 \Delta^{ }_{P-K}\Delta^{ }_{P-K-Q}} \biggr]_
              {p_0 = i E_p,Q = (2 i E_p,\vec{0})}
   \hspace*{-1cm}
 \;. \la{h11} 
\ea
After having expressed the integrals in these forms, we can make
use of Lorentz invariance in order to remove redundant structures: 
\ba
 \mbox{\nr{h4}} & = & \frac{E_p^2}{2M^2}
 \Bigl[
    \mbox{\nr{h2}} - \mbox{\nr{h1}}
 \Bigr]
 \;, \la{tra1} \\ 
 \mbox{\nr{h5}} & = & \frac{E_p^2}{2M^2}
 \Bigl[
    \mbox{\nr{h3}} - \mbox{\nr{h7}} - 2 M^2 \, \mbox{\nr{h10}}
 \Bigr]
 \;, \\[2mm] 
 \mbox{\nr{h9}} & = & 
    \mbox{\nr{h2}} - 2 M^2 \, \mbox{\nr{h7}} 
 \;, \\[2mm] 
 \mbox{\nr{h8}} & = & 
    \mbox{\nr{h2}} + 2(E_p^2 - M^2)\, \mbox{\nr{h6}} 
 \;. \la{tra4}
\ea
Inserting these relations, \eq\nr{coeff2} vanishes exactly. 
The coefficient of $D_{2E_p}(\tau)$, \eq\nr{coeff}, does not
vanish; after the transformation 
of \eqs\nr{tra1}--\nr{tra4} it can be written as
\ba
 & & \hspace*{-9mm} \mbox{``\eq\nr{coeff}''}  =  
 \nonumber \\[4mm] 
 & & \hspace*{-9mm}
 \biggl\{\int_K \mathbbm{P} \, \biggl[ 
   \frac{2(1-2\epsilon)}{M^2}\biggl( \frac{1}{K^2} - \frac{1}{\Delta^{ }_K}\biggr)
   - \frac{4 (1-\epsilon) }{K^2\Delta^{ }_{P-K}} 
   + \frac{2 (1 + \epsilon)}{\Delta_K^2}
     +  \biggl(3 + \frac{M^2}{E_p^2}\biggr) 
       \frac{2}{\Delta^{ }_K} \biggl( \frac{1}{\Delta^{ }_{K-Q}} -
               \frac{1}{\Delta^{ }_K} \biggr) 
 \nn & & 
   - \, \biggl(1 + \frac{M^2}{2E_p^2}\biggr) 
      \frac{8 M^2}{K^2 \Delta^2_{P-K}} 
   -  \biggl( 1 - \frac{M^4}{4 E_p^4} \biggr)
    \frac{16 E_p^2}{K^2\Delta^{ }_{P-K}\Delta^{ }_{P-K-Q}}
 \biggr]\biggr\}_{p_0 = i E_p,Q = (2 i E_p,\vec{0})} + \rmO(\epsilon)
 \nn & = & \frac{1}{4\pi^2} \biggl[ 
 \biggl( 3 + \frac{M^2}{E_p^2} \biggr)
 \biggl( 1 - \frac{p}{2 E_p} \ln\frac{E_p + p}{E_p - p}\biggr)
 - 1\biggr]  
 \nn & & \; + \,   
 \biggl(2  + \frac{M^2}{E_p^2} \biggr) \int_\vec{k} \, \mathbbm{P}
 \biggl\{ 
   \frac{M^2}
   { E_{pk} }
    \biggl[
      \frac{1}{2 \epsilon^{ }_k (\epsilon^{ }_k + E^{ }_p) \Delta_{++}^2}
     + \frac{1}{2 \epsilon^{ }_k (\epsilon^{ }_k - E^{ }_p) \Delta_{-+}^2}
     - \frac{1}{(\epsilon_k^2 - E_p^2 )E_{pk}^2} 
    \biggr]
 \nn & & \qquad\qquad\qquad + \, 
 \frac{2 E_p^2 - M^2}
 {2\epsilon_k^2 E^{ }_p E^{ }_{pk}}
    \biggl(
      \frac{1}{\Delta_{+}^{ }} 
     + \frac{1}{\Delta_{-}^{ }} 
     -  \frac{1}{\Delta_{++}^{ }} 
     +  \frac{1}{\Delta_{-+}^{ }} 
    \biggr)
 \biggr\} + \rmO(\epsilon)
 \;, 
\ea
where the first row represents the result of the ultraviolet 
sensitive integrals containing single or double propagators. 
The remaining $\vec{k}$-integral is infrared (IR) 
divergent and needs to be evaluated together with the other terms
of \eq\nr{NLO_tau}. Its form after angular integration 
can be found in \eq\nr{NLO_tau_final}, as the structure 
preceding $\theta(k)$.

It may be interesting to note that 
regulating the IR through a ``gluon mass'' $\lambda$,  
we obtain\footnote{%
 For \eq\nr{triangle2} a Feynman parameter $s$ can be used 
 for combining $\Delta^{ }_{P-K}$ and $\Delta^{ }_{P^*_{ }-K}$. 
 Subsequently the integral 
 $
 \{ \int_K \frac{1}{(K^2+\lambda^2) \Delta^{2}_{P-K}} \}^{ }_{P^2 = -M^2} = 
  \frac{1}{(4\pi)^2} \frac{1}{2M^2} \ln \frac{M^2}{\lambda^2} 
  + \rmO(\frac{\lambda^2}{M^4})
 $ can be used in both cases for determining the IR divergence. It is not
 enough for determining the term of $\rmO(1)$ in the latter case, 
 because the effective mass parameter appearing, 
 $\tilde M^2 = M^2 - 4 s(1-s)E_p^2$, crosses zero. Fortunately 
 the relevant
 integral over $s$ can be carried out without an expansion in $\lambda$.
 }
\ba
    \biggl\{\int_K \mathbbm{P} \, \biggl[
      \frac{8M^2}{(K^2 + \lambda^2) \Delta^2_{P-K}}  \biggr]
    \biggr\}_{p_0 = i E_p}
 \!\!\!\!   & = & 
 \frac{8}{(4\pi)^2} \ln\frac{M}{\lambda} + \rmO(\lambda)
 \;, \la{triangle1} \\
    \biggl\{\int_K \mathbbm{P} \, \biggl[
    \frac{16 E_p^2}{(K^2+\lambda^2)\Delta^{ }_{P-K}\Delta^{ }_{P^*_{ }-K}} \biggr]
    \biggr\}_{p_0 = i E_p}
 \!\!\!\!   & = & 
 \frac{8}{(4\pi)^2}
   \frac{E_p}{p} \biggl\{ \ln \biggl( \frac{E_p - p}{E_p + p} \biggr) 
   \ln\biggl( \frac{2p}{\lambda} \biggr) + \frac{\pi^2}{3} 
 \nn &  & \; +\, \mbox{Li}_2\biggl( \frac{E_p - p}{E_p + p} \biggr)
  + \fr14\! \ln^2 \biggl( \frac{E_p - p}{E_p + p} \biggr) 
   \biggr\} + \rmO(\lambda) \;. \la{triangle2}
 \nn
\ea
Here $P^*_{ }\equiv (-iE_p,\vec{p})$.
The sum of these (with the proper coefficients) 
amounts to a particular choice of 
$\int_0^\infty \! {\rm d}k \, \theta(k) / k$, 
appearing on the first row of \eq\nr{NLO_tau_final}; a corresponding
subtraction of $\theta(k)$ on the sixth row 
of \eq\nr{NLO_tau_final} would remove all 
powerlike terms at large $k$, with the price of inserting
more structure into the infrared regime $k\sim \lambda$. 

%

\end{document}